\documentclass[12pt]{article}

\usepackage{cite}
\usepackage{axodraw}
\usepackage{feynarts}

\makeatletter

\def\paragraph{\@startsection{paragraph}{4}{\z@}{+2.00ex plus
 +1ex minus +.2ex}{1.5ex plus .2ex}{\it\normalsize}}

\def\section{\@startsection {section}{1}{\z@}{+3.0ex plus +1ex minus
  +.2ex}{2.3ex plus .2ex}{\normalsize\bf\boldmath}}
\def\subsection{\@startsection{subsection}{2}{\z@}{+2.5ex plus +1ex
minus +.2ex}{1.5ex plus .2ex}{\normalsize\bf\boldmath}}
\def\subsubsection{\@startsection{subsubsection}{3}{\z@}{+3.25ex plus
 +1ex minus +.2ex}{1.5ex plus .2ex}{\normalsize\it}}

 \expandafter\ifx\csname mathrm\endcsname\relax\def\mathrm#1{{\rm #1}}\fi

\newcounter{saveeqn}

\@addtoreset{equation}{section}

\def\nl{\nonumber\\}

\newcommand{\lsim}
{\mathrel{\raisebox{-.3em}{$\stackrel{\displaystyle <}{\sim}$}}}
\newcommand{\gsim}
{\mathrel{\raisebox{-.3em}{$\stackrel{\displaystyle >}{\sim}$}}}
\def\asymp#1%
{\mathrel{\raisebox{-.4em}{$\widetilde{\scriptstyle #1}$}}}

\def\Nequal#1%
{\mathrel{\raisebox{-.5em}{$\stackrel{=}{\scriptstyle\rm#1}$}}}
\newcommand{\dsl}[1]{\not \hspace{-0.7mm}#1}
\def\dsl{\mathpalette\make@slash}
\def\make@slash#1#2{\setbox\z@\hbox{$#1#2$}%
  \hbox to 0pt{\hss$#1/$\hss\kern-\wd0}\box0}

\def\beq{\begin{equation}}
\def\eeq{\end{equation}}
\def\beqar{\begin{eqnarray}}
\def\eeqar{\end{eqnarray}}
\def\barr#1{\begin{array}{#1}}
\def\earr{\end{array}}
\def\bfi{\begin{figure}}
\def\efi{\end{figure}}
\def\btab{\begin{table}}
\def\etab{\end{table}}
\def\bce{\begin{center}}
\def\ece{\end{center}}
\def\nn{\nonumber}
\def\disp{\displaystyle}
\def\text{\textstyle}

\def\al{\alpha}

\def\Ga{\Gamma}
\def\ga{\gamma}
\def\de{\delta}

\def\la{\lambda}

\def\si{\sigma}

\def\reffi#1{\mbox{Figure~\ref{#1}}}
\def\reffis#1{\mbox{Figures~\ref{#1}}}
\def\refta#1{\mbox{Table~\ref{#1}}}

\def\refse#1{\mbox{Section~\ref{#1}}}

\def\citere#1{\mbox{Ref.~\cite{#1}}}
\def\citeres#1{\mbox{Refs.~\cite{#1}}}

\newcommand{\GeV}{\unskip\,\mathrm{GeV}}
\newcommand{\MeV}{\unskip\,\mathrm{MeV}}

\newcommand{\rd}{{\mathrm{d}}}

\newcommand{\Ord}{\mathswitch{{\cal{O}}}}
\newcommand{\Oa}{\mathswitch{{\cal{O}}(\alpha)}}

\newcommand{\Oas}{\mathswitch{{\cal{O}}(\alpha_s)}}
\newcommand{\A}{{\cal{A}}}

\newcommand{\M}{{\cal{M}}}

\def\mathswitchr#1{\relax\ifmmode{\mathrm{#1}}\else$\mathrm{#1}$\fi}

\newcommand{\PV}{V}
\newcommand{\PW}{\mathswitchr W}

\newcommand{\PZ}{\mathswitchr Z}

\newcommand{\Pg}{\mathswitchr g}

\newcommand{\PH}{\mathswitchr H}
\newcommand{\Pe}{\mathswitchr e}
\newcommand{\Pne}{\mathswitch \nu_{\mathrm{e}}}
\newcommand{\Pnebar}{\mathswitch \bar\nu_{\mathrm{e}}}

\newcommand{\Pd}{\mathswitchr d}
\newcommand{\Pdbar}{\bar{\mathswitchr d}}
\newcommand{\Pu}{\mathswitchr u}
\newcommand{\Pubar}{\bar{\mathswitchr u}}
\newcommand{\Ps}{\mathswitchr s}
\newcommand{\Psbar}{\bar{\mathswitchr s}}
\newcommand{\Pc}{\mathswitchr c}
\newcommand{\Pcbar}{\bar{\mathswitchr c}}
\newcommand{\Pb}{\mathswitchr b}

\newcommand{\Pt}{\mathswitchr t}

\newcommand{\Pep}{\mathswitchr {e^+}}
\newcommand{\Pem}{\mathswitchr {e^-}}

\def\mathswitch#1{\relax\ifmmode#1\else$#1$\fi}

\newcommand{\MV}{\mathswitch {M_\PV}}
\newcommand{\MW}{\mathswitch {M_\PW}}

\newcommand{\MZ}{\mathswitch {M_\PZ}}
\newcommand{\MH}{\mathswitch {M_\PH}}

\newcommand{\Mt}{\mathswitch {m_\Pt}}
\newcommand{\GW}{\Gamma_{\PW}}

\newcommand{\GZ}{\Gamma_{\PZ}}
\newcommand{\GV}{\Gamma_{\PV}}

\newcommand{\GF}{\mathswitch {G_\mu}}

\newcommand{\gs}{g_{\mathrm{s}}}
\newcommand{\als}{\al_{\mathrm{s}}}

\def\solid{\raise.9mm\hbox{\protect\rule{1.1cm}{.2mm}}}
\def\dash{\raise.9mm\hbox{\protect\rule{2mm}{.2mm}}\hspace*{1mm}}

\def\ie{i.e.\ }

\newcommand{\EW}{{\mathrm{EW}}}
\newcommand{\QCD}{{\mathrm{QCD}}}

\newcommand{\WW}{{\mathrm{WW}}}
\newcommand{\ZZ}{{\mathrm{ZZ}}}

\def\Re{\mathop{\mathrm{Re}}\nolimits}

\def\lra{\mathop{\mathrm{\leftrightarrow}}\nolimits}

\def\draftdate{\relax}
\def\mda{\relax}
\def\mua{\relax}
\def\mla{\relax}
\def\Mda{\relax}
\def\Mua{\relax}
\def\Mla{\relax}
\def\draft{
\def\thtystars{******************************}
\def\sixtystars{\thtystars\thtystars}
\typeout{}
\typeout{\sixtystars**}
\typeout{* Draft mode!
         For final version remove \protect\draft\space in source file *}
\typeout{\sixtystars**}
\typeout{}
\def\draftdate{\today}
\def\mua{\marginpar[\boldmath\hfil$\uparrow$]%
                   {\boldmath$\uparrow$\hfil}%
                    \typeout{marginpar: $\uparrow$}\ignorespaces}
\def\mda{\marginpar[\boldmath\hfil$\downarrow$]%
                   {\boldmath$\downarrow$\hfil}%
                    \typeout{marginpar: $\downarrow$}\ignorespaces}
\def\mla{\marginpar[\boldmath\hfil$\rightarrow$]%
                   {\boldmath$\leftarrow $\hfil}%
                    \typeout{marginpar: $\lra$}\ignorespaces}
\def\Mua{\marginpar[\boldmath\hfil$\Uparrow$]%
                   {\boldmath$\Uparrow$\hfil}%
                    \typeout{marginpar: $\uparrow$}\ignorespaces}
\def\Mda{\marginpar[\boldmath\hfil$\Downarrow$]%
                   {\boldmath$\Downarrow$\hfil}%
                    \typeout{marginpar: $\downarrow$}\ignorespaces}
\def\Mla{\marginpar[\boldmath\hfil$\Rightarrow$]%
                   {\boldmath$\Leftarrow $\hfil}%
                    \typeout{marginpar: $\lra$}\ignorespaces}
\overfullrule 5pt
\oddsidemargin -15mm
\marginparwidth 29mm
}

\def\stars{\strut\leaders\hbox{*}\hfill\strut}
\def\starline{\hfil\strut\hfil\hbox to \textwidth {\stars}\hfil}

\begin{document}
\enlargethispage{2cm}
\thispagestyle{empty}
\def\thefootnote{\fnsymbol{footnote}}
\setcounter{footnote}{1}
\null
\draftdate
\hfill KEK-CP-187\\
\strut\hfill KEK Preprint 2006-47\\
\strut\hfill MPP-2006-138\\
\strut\hfill PSI-PR-06-11\\
\strut\hfill WUB/06-08\\
\strut\hfill hep-ph/0611234\\
\begin{center}
{\Large \bf\boldmath
Radiative corrections to the 
\\[.1em]
semileptonic and hadronic Higgs-boson decays 
\\[.1em]
$\PH\to\PW\PW/\PZ\PZ\to4\,$fermions
\par} 
\vspace{.8cm}
{\large
{\sc A.\ Bredenstein$^1$, A.\ Denner$^2$, S.\ Dittmaier$^3$ 
and M.M.\ Weber$^4$} } \\[.5cm]
$^1$ {\it High Energy Accelerator Research
                Organization (KEK),
\\
Tsukuba, Ibaraki 305-0801, Japan} \\[0.5cm]
$^2$ {\it Paul Scherrer Institut, W\"urenlingen und Villigen,
\\
CH-5232 Villigen PSI, Switzerland} \\[0.5cm]
$^3$ {\it Max-Planck-Institut f\"ur Physik
(Werner-Heisenberg-Institut), \\
D-80805 M\"unchen, Germany}
\\[0.5cm]
$^4$ {\it Fachbereich Physik, Bergische Universit\"at Wuppertal,
\\
D-42097 Wuppertal, Germany}
\par \vskip 1em
\end{center}\par
\vfill {\bf Abstract:} \par The radiative corrections of the strong
and electroweak interactions are calculated for the Higgs-boson decays
$\PH\to\PW\PW/\PZ\PZ\to4f$ with semileptonic or hadronic four-fermion
final states in next-to-leading order.  This calculation is improved
by higher-order corrections originating from heavy-Higgs-boson effects
and photonic final-state radiation off charged leptons.  The W- and
Z-boson resonances are treated within the complex-mass scheme, i.e.\ 
without any resonance expansion or on-shell approximation.  The
calculation essentially follows our previous study of purely leptonic
final states.  The electroweak corrections are similar for all
four-fermion final states; for integrated quantities they amount to
some per cent and increase with growing Higgs-boson mass $\MH$,
reaching 7--8\% at $\MH\sim500\GeV$.  For distributions, the
corrections are somewhat larger and, in general, distort the shapes.
Among the QCD corrections, which include corrections to interference
contributions of the Born diagrams, only the corrections to the
squared Born diagrams turn out to be relevant. These contributions can
be attributed to the gauge-boson decays, i.e.\ they approximately
amount to $\als/\pi$ for semileptonic final states and $2\als/\pi$ for
hadronic final states.
The discussed corrections have been implemented in
the Monte Carlo event generator {\sc Prophecy4f}.%
\footnote{The computer code can be obtained from the authors upon request.}%
\par
\vskip .5cm
\noindent
November 2006 
\null
\setcounter{page}{0}
\clearpage
\def\thefootnote{\arabic{footnote}}
\setcounter{footnote}{0}

\section{Introduction}

The startup of the Large Hadron Collider (LHC) in 2007 will open up a
new era in particle physics. One of the main tasks of the LHC will be
the detection and the study of the Higgs boson.  If it is heavier than
$140\GeV$ and behaves as predicted by the Standard Model (SM), it
decays predominantly into gauge-boson pairs and subsequently into four
light fermions.  From a Higgs-boson mass $\MH$ of about $130\GeV$ up
to the Z-boson-pair threshold $2\MZ$, the decay signature
$\PH\to\PW\PW^*\to2\,$leptons + missing $p_{\mathrm{T}}$
\cite{Glover:1988fn} has the highest discovery potential for the Higgs
boson at the LHC \cite{Asai:2004ws}.  For higher Higgs-boson masses,
the leading role is taken over by the ``gold-plated'' channel
$\PH\to\PZ\PZ\to4\,$leptons, which will allow for the most accurate
measurement of $\MH$ above $130\GeV$ \cite{Zivkovic:2004sv}.  More
details and recent developments concerning Higgs-boson studies at the
LHC can be found in the literature \cite{atlas-tdr,cms-tdr}.  At a
future $\Pe^+\Pe^-$ linear collider \cite{Aguilar-Saavedra:2001rg},
the decays $\PH\to4f$ will enable measurements of the
$\PH\to\PW\PW/\PZ\PZ$ branching ratios at the level of a few to 10\%
\cite{Meyer:2004ha}.

At the LHC, owing to the huge background of strongly interacting
particles, the most important decay modes in
$\PH\to\PW\PW/\PZ\PZ\to4f$ are those with leptons in the final state.
Therefore, most analyses are based on them. However, also final states
involving quarks can be useful owing to their larger branching
fractions. For decays involving intermediate W bosons these provide
better kinematical information since they involve less neutrinos. For
instance, it has been found that in the vector-boson-fusion channel
the decays $\PH\to\PW\PW\to l^\pm\nu jj$ can provide complimentary
evidence in the intermediate Higgs-mass range $140\GeV<\MH<200\GeV$
\cite{Asai:2004ws,Cavasinni:2002,Pi:2006} and constitute a good potential
discovery channel in the medium-high Higgs-mass range
$\MH\gsim300\GeV$ \cite{atlas-tdr}. At a linear collider, the hadronic
and semileptonic final states are even more important since they allow
for a full reconstruction of the Higgs-boson decay $\PH\to\PW\PW$
\cite{Garcia-Abia:2005pf}.

A kinematical reconstruction of the Higgs-boson decays
$\PH\to\PW\PW\to4f$ and the suppression of the corresponding
backgrounds requires the study of distributions and the use of cuts
defined from the kinematics of the decay fermions. In addition, the
verification of the spin and of the CP properties of the Higgs boson
relies on the study of angular, energy, and invariant-mass
distributions \cite{Nelson:1986ki}.  These tasks require a Monte Carlo
generator for $\PH\to\PW\PW/\PZ\PZ\to4f$. Since the effects of
radiative corrections, in particular real-photon or gluon radiation,
are important, a Monte Carlo generator including all relevant
corrections is needed.

The progress in the theoretical description of the decays of a SM
Higgs boson into W- or Z-boson pairs has, for instance, been
summarized in \citere{Bredenstein:2006rh}. Until recently,
calculations for off-shell vector bosons were only available in lowest
order \cite{Djouadi:2005gi}, and radiative corrections were known only
in narrow-width approximation (NWA) \cite{Fleischer:1980ub}, i.e.\ for
on-shell W and Z~bosons.  In this case, also leading two-loop
corrections enhanced by powers of the top-quark mass
\cite{Kniehl:1995tn} or of the Higgs-boson mass
\cite{Ghinculov:1995bz} have been calculated.  However, near and below
the gauge-boson-pair thresholds the NWA is not applicable, so that
only the lowest-order results existed in this $\MH$ range.

In a recent paper \cite{Bredenstein:2006rh} we have presented results
for the complete electroweak (EW) ${\cal O}(\alpha)$ corrections
including some higher-order improvements to the Higgs-boson decays
$\PH\to\PW\PW/\PZ\PZ\to4\,$leptons.  First results of this calculation
had already been presented at the RADCOR05 conference
\cite{Bredenstein:2006nk}.  At this conference also progress on an
independent calculation of the electromagnetic corrections to
$\PH\to\PZ\PZ\to4\,$leptons has been reported by Carloni~Calame et
al.~\cite{CarloniCalame:2006vr}.  The analytic results demonstrated in
\citere{Bredenstein:2006rh} are also valid for quarks in the final
state. In this paper we supplement this calculation by the
corresponding QCD corrections. We introduce a classification of the
QCD corrections and describe their calculation.  The QCD corrections
have been implemented into the Monte Carlo generator {\sc Prophecy4f},
and numerical results have been produced. These include the partial
widths for various semileptonic and hadronic channels as well as
different invariant-mass and angular distributions for semileptonic
final states.

The paper is organized as follows: In \refse{se:setup} we describe the
setup of our calculation. Section \ref{se:QCDcorr} contains a
classification of the QCD corrections and provides analytic results
for the virtual and real QCD corrections. Numerical results are
presented in \refse{se:numerics}, and our conclusions are given
in \refse{se:concl}.

\section{Setup of the calculation}
\label{se:setup}

We consider the processes 
\beq\label{process-H4f}
\PH(p)
 \;\longrightarrow\;
f_1(k_1,\si_1) + \bar f_2(k_2,\si_2) + f_3(k_3,\si_3) + \bar
f_4(k_4,\si_4) + [\ga/\Pg(k,\la)],
\label{eq:h4f}
\eeq
where $f_i$ stands for any lepton,
$l=\Pe,\mu,\tau,\Pne,\nu_\mu,\nu_\tau$, or for any quark of the first
two generations, $q=\Pd,\Pu,\Ps,\Pc$.  We do not include final states
with bottom or top quarks.  The momenta and helicities of the external
particles are indicated in parentheses. The helicities take the values
$\sigma_i=\pm1/2$, but we often use only the sign to indicate the
helicity.  The masses of the external fermions are neglected whenever
possible, i.e.\ everywhere but in the mass-singular logarithms.  We
always sum over the four light quarks of the first two generations in
the final state and set the CKM matrix to the unit matrix. This
approximation ignores quark mixing with the third generation, which
is, however, negligible.

The calculation of the EW corrections has already been described in
\citere{Bredenstein:2006rh}, where results for purely leptonic final
states have been discussed.  Here we briefly repeat the salient
features of the evaluation of virtual one-loop and real-photonic
corrections.

The calculation of the one-loop diagrams has been performed in the
conventional 't~Hooft--Feynman gauge and in the background-field
formalism using the conventions of \citeres{Denner:1991kt} and
\cite{Denner:1994xt}, respectively.  

For the implementation of the finite widths of the gauge bosons we use
the complex-mass scheme, which was introduced in
\citere{Denner:1999gp} for lowest-order calculations and generalized
to the one-loop level in \citere{Denner:2005fg}.  In this approach the
W- and Z-boson masses are consistently considered as complex
quantities, defined as the locations of the propagator poles in the
complex plane.  The scheme fully respects all relations that follow
from gauge invariance.  A brief description of this scheme can also be
found in \citere{Denner:2006ic}.

The amplitudes have been generated with {\sc FeynArts}, using the two
independent versions 1 and 3, as described in
\citeres{Kublbeck:1990xc} and \cite{Hahn:2000kx}, respectively.  The
algebraic evaluation has been performed in two completely independent
ways. One calculation is based on an in-house program written in
{\sl Mathematica}, the other has been completed with the help of {\sc
  FormCalc} \cite{Hahn:1998yk}.  The amplitudes are expressed in terms
of standard matrix elements and coefficients, which contain the tensor
integrals, as described in the appendix of \citere{Denner:2003iy}.

The tensor integrals 
are evaluated as in the calculation of the
corrections to ${\rm e}^+{\rm e}^-\to4f$
\cite{Denner:2005fg,Denner:2005es}.  They are recursively reduced to
master integrals at the numerical level.  The scalar master integrals
are evaluated for complex masses using the methods and results of
\citere{'tHooft:1979xw}.  UV divergences are regulated dimensionally
and IR divergences with an infinitesimal photon mass.  Tensor and
scalar 5-point functions are directly expressed in terms of 4-point
integrals \cite{Denner:2002ii}.  Tensor 4-point and 3-point integrals
are reduced to scalar integrals with the Passarino--Veltman algorithm
\cite{Passarino:1979jh} as long as no small Gram determinant appears
in the reduction. If small Gram determinants occur, two alternative
schemes are applied \cite{Denner:2005nn}.  One method makes use of
expansions of the tensor coefficients about the limit of vanishing
Gram determinants and possibly other kinematical determinants.  In the
second, alternative method we evaluate a specific tensor coefficient,
the integrand of which is logarithmic in Feynman parametrization, by
numerical integration. Then the remaining coefficients as well as the
standard scalar integral are algebraically derived from this
coefficient.  The results of the two different codes, based on the
different methods described above are in good numerical agreement.

Since corrections due to the self-interaction of the Higgs boson
become important for large Higgs-boson masses, we have included the
dominant two-loop corrections to the decay ${\rm H}\to VV$
proportional to $G_\mu^2 M_{\rm H}^4$ in the large-Higgs-mass limit
which were calculated in \citere{Ghinculov:1995bz}.

The matrix elements for the real-photonic corrections are evaluated
using the Weyl--van der Waerden spinor technique as formulated in
\citere{Dittmaier:1998nn} and have been checked against results
obtained with {\sc Madgraph} \cite{Stelzer:1994ta}.  The soft and
collinear singularities are treated both in the dipole subtraction
method following \citere{Dittmaier:2000mb} and in the phase-space
slicing method following \citere{Bohm:1993qx}.  For the calculation of
non-collinear-safe observables we use the extension of the subtraction
method introduced in \citere{Bredenstein:2005zk}.  Final-state
radiation beyond ${\cal O}(\alpha)$ is included at the
leading-logarithmic level using the structure functions given in
\citere{Beenakker:1996kt} (see also references therein).

\begin{sloppypar}
  The phase-space integration is performed with Monte Carlo
  techniques.  {\sc Prophecy4f} employs a multi-channel Monte Carlo
  generator \cite{Berends:1994pv} similar to the one implemented in
  {\sc RacoonWW} \cite{Denner:1999gp} and {\sc Coffer}$\ga\ga$
  \cite{Bredenstein:2005zk,Bredenstein:2004ef}.  Our second code uses
  the adaptive integration program {\sc VEGAS} \cite{Lepage:1977sw}.
\end{sloppypar}

\section{QCD corrections for $\PH\to 2q2l$ and $\PH\to 4q$}
\label{se:QCDcorr}

\subsection{Classification}
\label{se:class}

A proper classification of QCD corrections is achieved upon
considering possible contributions to the squared lowest-order
amplitude.  The amplitude itself receives contributions from one of
the two tree diagrams shown in \reffi{fi:H4f-born-diags} or from both.
Thus, the square of this amplitude receives contributions from cut
diagrams of the types depicted in \reffi{fi:H4f-born-ints}.
\bfi
\centerline{
\setlength{\unitlength}{1pt}
\begin{picture}(190,100)(-20,0)
\DashLine(15,50)(60,50){3}
\Photon(60,50)(90,20){-2}{5}
\Photon(60,50)(90,80){2}{5}
\Vertex(60,50){2.0}
\Vertex(90,80){2.0}
\Vertex(90,20){2.0}
\ArrowLine(90,80)(120, 95)
\ArrowLine(120,65)(90,80)
\ArrowLine(120, 5)( 90,20)
\ArrowLine( 90,20)(120,35)
\put(0,47){$\PH$}
\put(62,70){$V$}
\put(62,18){$V$}
\put(125,90){$f_a$}
\put(125,65){$\bar f_b$}
\put(125,30){$f_c$}
\put(125,5){$\bar f_d$}
\end{picture}
\begin{picture}(190,100)(-20,0)
\DashLine(15,50)(60,50){3}
\Photon(60,50)(90,35){-2}{4}
\Photon(60,50)(90,80){2}{5}
\Vertex(60,50){2.0}
\Vertex(90,80){2.0}
\Vertex(90,35){2.0}
\ArrowLine(90,80)(120, 95)
\ArrowLine(120,65)(90,35)
\ArrowLine(120, 5)(90,80)
\ArrowLine( 90,35)(120,35)
\put(0,47){$\PH$}
\put(62,70){$V'$}
\put(62,18){$V'$}
\put(125,90){$f_a$}
\put(125,65){$\bar f_b$}
\put(125,30){$f_c$}
\put(125,5){$\bar f_d$}
\end{picture}
}
\caption{Possible lowest-order diagrams for $\PH\to4f$ where $V,V'=\PW,\PZ$.}
\label{fi:H4f-born-diags}
\vspace*{3em}
\centerline{
\setlength{\unitlength}{1pt}
\begin{picture}(180,100)
\DashLine( 0,50)(30,50){3}
\Photon(30,50)(60,20){-2}{5}
\Photon(30,50)(60,80){2}{5}
\Vertex(30,50){2.0}
\Vertex(60,80){2.0}
\Vertex(60,20){2.0}
\ArrowLine(60,80)(90,95)
\ArrowLine(90,65)(60,80)
\ArrowLine(90, 5)(60,20)
\ArrowLine(60,20)(90,35)
\DashLine(90,0)(90,100){7}
\ArrowLine(90,95)(120,80)
\ArrowLine(120,80)(90,65)
\ArrowLine(120,20)(90, 5)
\ArrowLine(90,35)(120,20)
\Vertex(150,50){2.0}
\Vertex(120,80){2.0}
\Vertex(120,20){2.0}
\Photon(150,50)(120,20){2}{5}
\Photon(150,50)(120,80){-2}{5}
\DashLine(180,50)(150,50){3}
\put(0,90){(A)}
\end{picture}
\hspace*{2em}
\begin{picture}(180,100)
\DashLine(0,50)(30,50){3}
\Photon(30,50)(60,35){-2}{4}
\Photon(30,50)(60,80){2}{5}
\Vertex(30,50){2.0}
\Vertex(60,80){2.0}
\Vertex(60,35){2.0}
\ArrowLine(60,80)(90,95)
\ArrowLine(90,65)(60,35)
\ArrowLine(90, 5)(60,80)
\ArrowLine(60,35)(90,35)
\DashLine(90,0)(90,100){7}
\ArrowLine(90,95)(120,80)
\ArrowLine(120,80)(90,65)
\ArrowLine(120,20)(90, 5)
\ArrowLine(90,35)(120,20)
\Vertex(150,50){2.0}
\Vertex(120,80){2.0}
\Vertex(120,20){2.0}
\Photon(150,50)(120,20){2}{5}
\Photon(150,50)(120,80){-2}{5}
\DashLine(180,50)(150,50){3}
\put(0,90){(B)}
\end{picture}
}
\caption{Types of cut diagrams contributing in lowest order.}
\label{fi:H4f-born-ints}
\efi
Type (A) corresponds to the squares of each of the Born diagrams,
type (B) to their interference if two Born diagrams exist.

After this preliminary consideration we define four different
categories of QCD corrections. Examples of cut diagrams belonging to
these categories are shown in \reffi{fi:virtual-ints}, the corresponding
virtual QCD correction diagrams are depicted in \reffi{fi:virtual}.
\begin{enumerate}
\renewcommand{\labelenumi}{(\alph{enumi})}
\item {\it QCD corrections to gauge-boson decays} comprise all cut
  diagrams resulting from diagram (A) of \reffi{fi:H4f-born-ints} by
  adding one additional gluon.  Cut diagrams in which the gluon does
  not cross the cut correspond to virtual one-loop corrections, the
  one where the gluon crosses the cut correspond to real-gluon
  radiation. Note that cut diagrams in which the gluon connects the
  two closed quark lines identically vanish, because their colour
  structure is proportional to
  $\mathrm{Tr}(\la^h)\mathrm{Tr}(\la^h)=0$, where $\la^h$ is a
  Gell-Mann matrix.  Thus, the only relevant one-loop diagrams in this
  category are gluonic vertex corrections to a weak-boson decay, as
  illustrated in the first diagram of \reffi{fi:virtual}; the real
  corrections are induced by the corresponding gluon bremsstrahlung
  diagrams.
\bfi
\begin{center}
{\setlength{\unitlength}{.9pt}\SetScale{.9}
\begin{picture}(210,100)
\DashLine( 0,50)(30,50){3}
\Photon(30,50)(60,20){-2}{5}
\Photon(30,50)(60,80){2}{5}
\Vertex(30,50){2.0}
\Vertex(60,80){2.0}
\Vertex(60,20){2.0}
\ArrowLine(60,80)(90,95)
\ArrowLine(90,95)(120,95)
\ArrowLine(120,65)(90,65)
\ArrowLine(90,65)(60,80)
\Vertex(90,95){2.0}
\Vertex(90,65){2.0}
\Gluon(90,65)(90,95){2}{4}
\SetColor{Black}
\ArrowLine(120, 5)(60,20)
\ArrowLine(60,20)(120,35)
\DashLine(120,0)(120,100){7}
\ArrowLine(120,95)(150,80)
\ArrowLine(150,80)(120,65)
\ArrowLine(150,20)(120, 5)
\ArrowLine(120,35)(150,20)
\Vertex(180,50){2.0}
\Vertex(150,80){2.0}
\Vertex(150,20){2.0}
\Photon(180,50)(150,20){2}{5}
\Photon(180,50)(150,80){-2}{5}
\DashLine(210,50)(180,50){3}
\put(0,90){{(a)}}
\end{picture}
\hspace*{2em}
\begin{picture}(210,100)
\DashLine(210,50)(180,50){3}
\Photon(180,50)(160,35){-2}{4}
\Photon(180,50)(160,80){2}{5}
\Vertex(180,50){2.0}
\Vertex(160,80){2.0}
\Vertex(160,35){2.0}
\ArrowLine(120,95)(160,80)
\ArrowLine(160,35)(120,65)
\ArrowLine(160,80)(120, 5)
\ArrowLine(120,35)(160,35)
\DashLine(120,0)(120,100){7}
\ArrowLine(60,80)(120,95)
\ArrowLine(90,65)(60,80)
\ArrowLine(120,65)(90,65)
\ArrowLine(120, 5)(60,20)
\ArrowLine(60,20)(90,35)
\ArrowLine(90,35)(120,35)
\Vertex(90,35){2.0}
\Vertex(90,65){2.0}
\Gluon(90,35)(90,65){2}{4}
\SetColor{Black}
\Vertex(30,50){2.0}
\Vertex(60,80){2.0}
\Vertex(60,20){2.0}
\Photon(30,50)(60,20){2}{5}
\Photon(30,50)(60,80){-2}{5}
\DashLine(0,50)(30,50){3}
\put(0,90){{(b)}}
\end{picture}
\\[2em]
\begin{picture}(200,100)(15,0)
\ArrowLine(120,95)(150,80)
\ArrowLine(150,80)(120,65)
\ArrowLine(150,20)(120, 5)
\ArrowLine(120,35)(150,20)
\Vertex(180,50){2.0}
\Vertex(150,80){2.0}
\Vertex(150,20){2.0}
\Photon(180,50)(150,20){2}{5}
\Photon(180,50)(150,80){-2}{5}
\DashLine(210,50)(180,50){3}
\DashLine(120,0)(120,100){7}
\DashLine(15,50)(50,50){3}
\Photon(50,50)(70,35){-2}{3}
\Photon(50,50)(70,65){2}{3}
\Vertex(50,50){2.0}
\Vertex(70,35){2.0}
\Vertex(70,65){2.0}
\ArrowLine(70,65)(120, 95)
\ArrowLine(70,35)(85,50)
\ArrowLine(85,50)(70,65)
\ArrowLine(120, 5)(70,35)
\ArrowLine(120,65)(105,50)
\ArrowLine(105,50)(120,35)
\Vertex(85,50){2.0}
\Gluon(85,50)(105,50){2}{3}
\Vertex(105,50){2.0}
\SetColor{Black}
\put(8,90){{(c)}}
\end{picture}
\hspace*{2em}
\begin{picture}(200,100)(8,0)
\DashLine(210,50)(180,50){3}
\Photon(180,50)(160,35){-2}{4}
\Photon(180,50)(160,80){2}{5}
\Vertex(180,50){2.0}
\Vertex(160,80){2.0}
\Vertex(160,35){2.0}
\ArrowLine(120,95)(160,80)
\ArrowLine(160,35)(120,65)
\ArrowLine(160,80)(120, 5)
\ArrowLine(120,35)(160,35)
\DashLine(120,0)(120,100){7}
\DashLine(15,50)(50,50){3}
\ArrowLine(70,35)(50,50)
\ArrowLine(50,50)(70,65)
\ArrowLine(70,65)(70,35)
\Vertex(50,50){2.0}
\ArrowLine(95,80)(120,95)
\ArrowLine(120,65)(95,80)
\ArrowLine(95,20)(120,35)
\ArrowLine(120, 5)(95,20)
\Vertex(70,35){2.0}
\Vertex(70,65){2.0}
\Gluon(70,65)(95,80){2}{4}
\Gluon(70,35)(95,20){-2}{4}
\Vertex(95,80){2.0}
\Vertex(95,20){2.0}
\SetColor{Black}
\put(8,90){{(d)}}
\end{picture}
}
\vspace*{-1em}
\end{center}
\caption{Categories of cut diagrams contributing to the QCD corrections.}
\label{fi:virtual-ints}
\efi

\bfi
\begin{center}
\setlength{\unitlength}{1pt}
\begin{picture}(190,100)(-20,10)
\DashLine(15,50)(60,50){3}
\Photon(60,50)(90,20){-2}{5}
\Photon(60,50)(90,80){2}{5}
\Vertex(60,50){2.0}
\Vertex(90,80){2.0}
\Vertex(90,20){2.0}
\ArrowLine(120, 5)( 90,20)
\ArrowLine( 90,20)(120,35)
\ArrowLine( 90,80)(110,90)
\ArrowLine(110,90)(120,95)
\ArrowLine(120,65)(110,70)
\ArrowLine(110,70)(90,80)
\Vertex(110,90){2.0}
\Vertex(110,70){2.0}
\Gluon(110,90)(110,70){2}{3}
\put(-6,47){$\PH$}
\put(62,70){$V$}
\put(62,18){$V$}
\put(125,90){$f_a$}
\put(125,65){$\bar f_b$}
\put(125,30){$f_c$}
\put(125,5){$\bar f_d$}
\put(117,77.5){$\Pg$}
\put(-22,100){(a,b)}
\end{picture}
\begin{picture}(190,100)(-20,10)
\DashLine(15,50)(60,50){3}
\Photon(60,50)(90,20){-2}{5}
\Photon(60,50)(90,80){2}{5}
\Vertex(60,50){2.0}
\Vertex(90,80){2.0}
\Vertex(90,20){2.0}
\ArrowLine(90,80)(120, 95)
\ArrowLine(120,65)(105,72.5)
\ArrowLine(105,72.5)(90,80)
\ArrowLine( 90,20)(105,27.5)
\ArrowLine(105,27.5)(120,35)
\ArrowLine(120, 5)( 90,20)
\Vertex(105,72.5){2.0}
\Vertex(105,27.5){2.0}
\Gluon(105,72.5)(105,27.5){2}{8}
\put(-6,47){$\PH$}
\put(62,70){$V$}
\put(62,18){$V$}
\put(125,90){$f_a$}
\put(125,65){$\bar f_b$}
\put(125,30){$f_c$}
\put(125,5){$\bar f_d$}
\put(112,47.5){$\Pg$}
\put(-22,100){(b)}
\end{picture}\\[2em]
\begin{picture}(190,100)(-20,10)
\DashLine(15,50)(50,50){3}
\Photon(50,50)(70,35){-2}{3}
\Photon(50,50)(70,65){2}{3}
\Vertex(50,50){2.0}
\Vertex(70,35){2.0}
\Vertex(70,65){2.0}
\Vertex(85,50){2.0}
\ArrowLine(70,65)(120, 95)
\ArrowLine(70,35)(85,50)
\ArrowLine(85,50)(70,65)
\ArrowLine(120, 5)(70,35)
\Gluon(85,50)(105,50){2}{3}
\Vertex(105,50){2.0}
\ArrowLine(105,50)(120,65)
\ArrowLine(120,35)(105,50)
\put(-6,47){$\PH$}
\put(52,66){$V$}
\put(52,22){$V$}
\put(125,90){$f_a$}
\put(125,65){$f_c$}
\put(125,30){$\bar f_d$}
\put(125,5){$\bar f_b$}
\put(92,38){$\Pg$}
\put(-22,100){(c)}
\end{picture}
\begin{picture}(190,100)(-20,10)
\DashLine(15,50)(50,50){3}
\ArrowLine(70,35)(50,50)
\ArrowLine(50,50)(70,65)
\ArrowLine(70,65)(70,35)
\Vertex(50,50){2.0}
\Vertex(70,35){2.0}
\Vertex(70,65){2.0}
\Gluon(70,65)(95,80){2}{4}
\Gluon(70,35)(95,20){-2}{4}
\Vertex(95,80){2.0}
\Vertex(95,20){2.0}
\ArrowLine(95,80)(120,95)
\ArrowLine(120,65)(95,80)
\ArrowLine(95,20)(120,35)
\ArrowLine(120, 5)(95,20)
\put(-6,47){$\PH$}
\put(54,66){$q$}
\put(54,27){$q$}
\put(76,47){$q$}
\put(125,90){$f_a$}
\put(125,65){$\bar f_b$}
\put(125,30){$f_c$}
\put(125,5){$\bar f_d$}
\put(78,15){$\Pg$}
\put(78,82){$\Pg$}
\put(-22,100){(d)}
\end{picture}
\end{center}
\caption{Basic diagrams contributing to the virtual QCD corrections for
  $\PH\to4f$ where $V=\PW,\PZ$ and $q=\Pd,\Pu,\Ps,\Pc,\Pb,\Pt$.  The
  categories of QCD corrections, (a)--(d), to which the diagrams
  contribute are indicated.}
\label{fi:virtual}
\efi

If a weak-boson decay is fully integrated over its decay angles, the
resulting QCD correction of the considered type simply reduces to the
well-known factor $\als/\pi$ for a hadronically decaying vector boson.
\item {\it QCD corrections to interferences} comprise all cut diagrams
  resulting from diagram (B) of \reffi{fi:H4f-born-ints} by adding one
  additional gluon, analogously to the previous category.  Relevant
  one-loop diagrams are, thus, vertex corrections or pentagon
  diagrams, as illustrated in the first two diagrams of
  \reffi{fi:virtual}.
\item {\it Corrections from intermediate $q\bar q\Pg^*$ states} are
  induced by loop diagrams exemplified by the third graph in
  \reffi{fi:virtual}. The remaining graphs are obtained by shifting
  the gluon to different positions at the same quark line and by
  interchanging the role of the two quark lines.  Thus, the diagrams
  comprise not only box diagrams but also vertex diagrams.  They do
  not interfere with Born diagrams with the same fermion-number flow
  because of the colour structure, i.e.\ in $\Ord(\als)$ they only
  contribute if two Born diagrams exist.
  
  Owing to the intermediate $q\bar q\Pg^*$ states, the squared
  diagrams of this category actually correspond to
  (collinear-singular) real NLO QCD corrections to the loop-induced
  decay $\PH\to q\bar q\Pg$, where $q$ is a massless quark. Here we
  consider only the interference contributions of the loop diagrams of
  this category with the lowest-order diagrams for the decay $\PH\to
  VV\to4q$, resulting in a UV and IR (soft and collinear) finite
  correction.
\item {\it Corrections from intermediate $\Pg^*\Pg^*$ states} are
  induced by diagrams exemplified by the fourth graph in
  \reffi{fi:virtual}. There are precisely two graphs with opposite
  fermion-number flow in the loop. Again, owing to the colour
  structure (see also below), these diagrams do not interfere with
  Born diagrams with the same fermion-number flow, i.e.\ the existence
  of two Born diagrams is needed.
  
  Owing to the intermediate $\Pg^*\Pg^*$ states, the squared diagrams
  of this category actually correspond to (collinear-singular) real
  NNLO QCD corrections to the loop-induced decay $\PH\to\Pg\Pg$.  The
  considered interference contributions of the loop diagrams of this
  category with the lowest-order diagrams for the decay $\PH\to
  VV\to4q$, however, again yield a UV and IR (soft and collinear)
  finite correction.
\end{enumerate}

From the classification, it is clear that category (a) exists for all
final states involving quarks, while categories (b), (c), and (d) are
only relevant for the hadronic decays $\PH\to q\bar q q\bar q$ and
$\PH\to q\bar q q'\bar q'$, where $q$ and $q'$ are weak-isospin
partners.  Categories (a), (b), and (c) give rise to contributions
to the decay widths that are
proportional to $\alpha^3\als$, while type (d) yields a contribution
proportional to $\alpha^2\als^2$.

We do not consider the process $\PH\to4\,\mathrm{jets}$ in general but
only the contributions via virtual EW gauge-boson pairs, i.e.\ we
assume that the gauge-boson resonances are isolated by experimental
cuts.  For the more inclusive decay $\PH\to4\,\mathrm{jets}$, also
diagrams without intermediate EW gauge bosons, where the Higgs boson
couples to gluons via heavy-quark loops, become important.  Using an
effective $\PH\Pg\Pg$ coupling, the calculation of the corresponding
QCD one-loop matrix elements has been described in
\citere{Ellis:2005qe}, but the full NLO QCD prediction for
$\PH\to4\,\mathrm{jets}$ including these effects is not yet available.

\subsection{Virtual corrections}
\label{se:virtcorr}

In the evaluation of the one-loop QCD diagrams, which are illustrated
in \reffi{fi:virtual}, the fermion spinor chains are separated from
the rest of the amplitude by introducing 52 standard matrix elements 
$\hat\M^{abcd,\si\tau}_i$, as defined in Eq.~(3.2) of
\citere{Bredenstein:2006rh}, where the indices $\si$ and
$\tau$ indicate the chiralities in the spinor chains of the fermion
pairs $f_a\bar f_b$ and $f_c\bar f_d$, respectively.
Furthermore, the colour structure is extracted by defining the
colour operators
\beq
C_1^{abcd}=\de_{c_ac_b}\otimes\de_{c_cc_d}, \qquad
C_2^{abcd}=\frac{1}{4C_{\mathrm{F}}}\la^h_{c_ac_b}\otimes\la^h_{c_cc_d} =
\frac{3}{16}\la^h_{c_ac_b}\otimes\la^h_{c_cc_d}
\eeq
with the Gell-Mann matrices $\la^h$, the colour index $h$ of the gluon,
and the colour indices $c_{a,b,c,d}$ of the quarks.  For external
leptons the corresponding colour index trivially takes only one value,
and the operator $C_2$, of course, appears only for four-quark final
states.  Using this notation, the generic lowest-order amplitude in
colour space reads
\beq
\A_{0,c_a c_b c_c c_d}^{VV,\si_a\si_b\si_c\si_d}(k_a,k_b,k_c,k_d) =
C_1^{abcd}\M_0^{VV,\si_a\si_b\si_c\si_d}(k_a,k_b,k_c,k_d),
\eeq
\begin{sloppypar}
\noindent
where $\M_0^{VV,\si_a\si_b\si_c\si_d}$ is the colour-stripped
generic lowest-order amplitude defined in Eq.~(2.7) of
\citere{Bredenstein:2006rh}.
Obviously, this notation generalizes to the generic 
EW one-loop amplitudes (i.e.\ without gluon exchange) introduced in
Eq.~(3.3) of \citere{Bredenstein:2006rh},%
\footnote{We note that the generic colour-stripped EW one-loop 
amplitude $\M^{VV,\si_a\si_b\si_c\si_d}_{\EW}$ was denoted 
$\M^{abcd,\si\tau}_1$ in Eq.~(3.3) of \citere{Bredenstein:2006rh}.}
\end{sloppypar}
\newcommand{\Msme}{\hat\M}
\beq
\A^{VV,\si_a\si_b\si_c\si_d}_{\EW,c_a c_b c_c c_d} =
C_1^{abcd}\M^{VV,\si_a\si_b\si_c\si_d}_{\EW} =
C_1^{abcd}\sum_{i=1}^{13} F^{abcd,\si_a\si_c}_{\EW,i} 
\Msme^{abcd,\si_a\si_c}_i \,\delta_{\si_a,-\si_b} \delta_{\si_c,-\si_d},
\eeq
where $\Msme^{abcd,\si_a\si_c}_i$ denote the standard matrix elements
and $F^{abcd,\si_a\si_c}_{\EW,i}$ are Lorentz-invariant coefficient
functions.  In the generic amplitudes the superscript ``$VV$''
indicates the common fermion-number flow, which corresponds to the
decays $\PV\to f_a\bar f_b$ and $\PV\to f_c\bar f_d$.  The one-loop
QCD amplitude, which involves gluon exchange, receives contributions
from both colour operators; in colour space we define
\beqar
\A_{\QCD,c_a c_b c_c c_d}^{VV,\si_a\si_b\si_c\si_d} &=&
\sum_{j=1}^2C_j^{abcd}\M^{VV,\si_a\si_b\si_c\si_d}_{\QCD,j},
\nn\\ 
\M^{VV,\si_a\si_b\si_c\si_d}_{\QCD,j} &=&
\sum_{i=1}^{13}F_{\QCD,ji}^{abcd,\si_a\si_c} \Msme^{abcd,\si_a\si_c}_i
\,\delta_{\si_a,-\si_b} \delta_{\si_c,-\si_d},
\eeqar
where the $\M^{VV,\si_a\si_b\si_c\si_d}_{\QCD,j}$ are colour-stripped 
amplitudes.

From the generic matrix elements $\A_{n,c_a c_b c_c
  c_d}^{VV,\si_a\si_b\si_c\si_d}$ ($n=0,1$) the matrix elements
$\A_{n,c_a c_b c_c c_d}^{\si_a\si_b\si_c\si_d}$ for the specific
processes are constructed as in Eqs.~(2.11)--(2.14) of
\citere{Bredenstein:2006rh}.  The index $n=1$ collectively represents
the sum $\EW+\QCD$ of EW and QCD one-loop contributions.  We denote
different fermions by $f$ and $F$, and their weak-isospin partners by
$f'$ and $F'$ ($f\ne F,F'$). For purely hadronic final states the
quarks are denoted by $q$ and their weak-isospin partners by $q'$.
Thus, we obtain:
\begin{itemize}
\item $\PH\to f\bar f F\bar F$:
\beqar
\A_{n,c_1 c_2 c_3 c_4}^{\si_1\si_2\si_3\si_4}(k_1,k_2,k_3,k_4)&=&
\A_{n,c_1 c_2 c_3 c_4}^{\ZZ,\si_1\si_2\si_3\si_4}(k_1,k_2,k_3,k_4),
\eeqar
\item $\PH\to f\bar f' F\bar F'$:
\beqar
\A_{n,c_1 c_2 c_3 c_4}^{\si_1\si_2\si_3\si_4}(k_1,k_2,k_3,k_4)
&=&\A_{n,c_1 c_2 c_3 c_4}^{\WW,\si_1\si_2\si_3\si_4}(k_1,k_2,k_3,k_4),
\eeqar
\item $\PH\to q\bar q q\bar q$:
\beqar
\A_{n,c_1 c_2 c_3 c_4}^{\si_1\si_2\si_3\si_4}(k_1,k_2,k_3,k_4)&=&
\A_{n,c_1 c_2 c_3 c_4}^{\ZZ,\si_1\si_2\si_3\si_4}(k_1,k_2,k_3,k_4)
\nl&&{}
-\A_{n,c_1 c_4 c_3 c_2}^{\ZZ,\si_1\si_4\si_3\si_2}(k_1,k_4,k_3,k_2),
\eeqar
\item $\PH\to q\bar q q'\bar q'$:
\beqar
\A_{n,c_1 c_2 c_3 c_4}^{\si_1\si_2\si_3\si_4}(k_1,k_2,k_3,k_4)
&=&\A_{n,c_1 c_2 c_3 c_4}^{\ZZ,\si_1\si_2\si_3\si_4}(k_1,k_2,k_3,k_4)
\nl&&{}
-\A_{n,c_1 c_4 c_3 c_2}^{\WW,\si_1\si_4\si_3\si_2}(k_1,k_4,k_3,k_2).
\eeqar
\end{itemize}
The relative signs between contributions of the basic subamplitudes to
the full matrix elements account for the sign changes resulting from
interchanging external fermion lines.

Since the lowest-order amplitudes only involve the colour operators
$C^{1234}_1$ and $C^{1432}_1$, the following colour sums appear in the
calculation of squared lowest-order amplitudes and of interferences
between one-loop and lowest-order matrix elements:
\beq
\begin{array}[b]{rclcrcl}
X^{(A)}_1 &=& \disp\sum_{\{c_i\}}(C^{abcd\,*}_1C^{abcd}_1) =
N^{\mathrm{c}}_{f_a} N^{\mathrm{c}}_{f_c},
&\qquad& X^{(A)}_2 &=& \disp\sum_{\{c_i\}}(C^{abcd\,*}_1C^{abcd}_2) =0,
\\[1.3em]
X^{(B)}_1 &=& \disp\sum_{\{c_i\}}(C^{abcd\,*}_1C^{adcb}_1) =
N^{\mathrm{c}}_{f_a},
&& X^{(B)}_2 &=& \disp\sum_{\{c_i\}}(C^{abcd\,*}_1C^{adcb}_2) =
N^{\mathrm{c}}_{f_a},
\end{array}
\eeq
where $\sum_{\{c_i\}}$ stands for the sum over the colour indices
$c_a$, $c_b$, $c_c$, $c_d$, and $N^{\mathrm{c}}_{f}$ is the colour
factor for a fermion $f$, which is 1 for leptons and 3 for quarks.

Squared Born diagrams, as illustrated in type (A) of
\reffi{fi:H4f-born-diags}, are proportional to $X^{(A)}_1$,
lowest-order interference diagrams of type (B) are proportional to
$X^{(B)}_1$.  The situation is analogous for all one-loop diagrams
without gluons.  By definition, category (a) of the gluonic diagrams
comprises all one-loop QCD corrections proportional to $X^{(A)}_1$.
In category (b), the vertex corrections are proportional to
$X^{(B)}_1$ and the pentagons to $X^{(B)}_2$.  Categories (c) and (d)
receive only contributions from $X^{(B)}_2$; interferences of one-loop
diagrams like (c) and (d) in \reffi{fi:virtual} with Born diagrams of
the same fermion-number flow vanish because of $X^{(A)}_2=0$.

Finally, we obtain the following for the one-loop corrections to the
squared matrix elements:
\begin{itemize}
\item $\PH\to f\bar f F\bar F$:
\beqar
\sum_{\{c_i\}} 2\Re\left\{
\A_{0,c_1 c_2 c_3 c_4}^{\si_1\si_2\si_3\si_4\,*}
\A_{1,c_1 c_2 c_3 c_4}^{\si_1\si_2\si_3\si_4} \right\}
&=&
2\Re\left\{
N^{\mathrm{c}}_{f} N^{\mathrm{c}}_{F}
\M_{0}^{\PZ\PZ,\si_1\si_2\si_3\si_4\,*}
\M_{\EW+\QCD\mathrm{(a)}}^{\PZ\PZ,\si_1\si_2\si_3\si_4} \right\},
\hspace{3em}
\eeqar
\item $\PH\to f\bar f' F\bar F'$:
\beqar
\sum_{\{c_i\}} 2\Re\left\{
\A_{0,c_1 c_2 c_3 c_4}^{\si_1\si_2\si_3\si_4\,*}
\A_{1,c_1 c_2 c_3 c_4}^{\si_1\si_2\si_3\si_4} \right\}
&=&
2\Re\left\{
N^{\mathrm{c}}_{f} N^{\mathrm{c}}_{F}
\M_{0}^{\PW\PW,\si_1\si_2\si_3\si_4\,*}
\M_{\EW+\QCD\mathrm{(a)}}^{\PW\PW,\si_1\si_2\si_3\si_4} \right\},
\hspace{3em}
\eeqar
\item $\PH\to q\bar q q\bar q$:
\beqar
\lefteqn{
\sum_{\{c_i\}} 2\Re\left\{
\A_{0,c_1 c_2 c_3 c_4}^{\si_1\si_2\si_3\si_4\,*}
\A_{1,c_1 c_2 c_3 c_4}^{\si_1\si_2\si_3\si_4} \right\}
} \hspace*{1em}
&&
\nn\\
&=&
2\Re\Bigl\{
\M_{0}^{\PZ\PZ,\si_1\si_2\si_3\si_4\,*} \Bigl[
(N^{\mathrm{c}}_{q})^2 \,
\M_{\EW+\QCD\mathrm{(a)}}^{\PZ\PZ,\si_1\si_2\si_3\si_4} 
-N^{\mathrm{c}}_{q} \,
\M_{\EW+\QCD(\mathrm{b})+\QCD\mathrm{(c)}+\QCD\mathrm{(d)}}^{\PZ\PZ,\si_1\si_4\si_3\si_2}
\Bigr]
\Bigr\}
\nn\\
&& 
+\; \Bigl(q(k_2,\si_2)\leftrightarrow q(k_4,\si_4)\Bigr),
\eeqar
\item $\PH\to q\bar q q'\bar q'$:
\beqar
\lefteqn{
\sum_{\{c_i\}} 2\Re\left\{
\A_{0,c_1 c_2 c_3 c_4}^{\si_1\si_2\si_3\si_4\,*}
\A_{1,c_1 c_2 c_3 c_4}^{\si_1\si_2\si_3\si_4} \right\}
} \hspace*{1em}
&&
\nn\\
&=&
2\Re\Bigl\{\phantom{{}+{}}
\M_{0}^{\PZ\PZ,\si_1\si_2\si_3\si_4\,*} \Bigl[
(N^{\mathrm{c}}_{q})^2 \,
\M_{\EW+\QCD\mathrm{(a)}}^{\PZ\PZ,\si_1\si_2\si_3\si_4} 
-N^{\mathrm{c}}_{q} \,
\M_{\EW+\QCD(\mathrm{b})}^{\PW\PW,\si_1\si_4\si_3\si_2}
\Bigr]
\nn\\
&& \phantom{2\Re\Bigl\{}
+\M_{0}^{\PW\PW,\si_1\si_4\si_3\si_2\,*} \Bigl[
\phantom{{}-{}}
(N^{\mathrm{c}}_{q})^2 \,
\M_{\EW+\QCD\mathrm{(a)}}^{\PW\PW,\si_1\si_4\si_3\si_2} 
\nn\\
&& \hspace*{10.7em}
{}-N^{\mathrm{c}}_{q} \,
\M_{\EW+\QCD(\mathrm{b})+\QCD\mathrm{(c)}+\QCD\mathrm{(d)}}^{\PZ\PZ,\si_1\si_2\si_3\si_4}
\Bigr]
\Bigr\}.
\eeqar
\end{itemize}
Due to the electric charge flow, categories (c) and (d) only exist if
there are corresponding diagrams with intermediate Z bosons. That is
why there are no terms $\M_{0}^{\PZ\PZ,\si_1\si_2\si_3\si_4\,*}\times
\M_{\QCD(\mathrm{c})+\QCD(\mathrm{d})}^{\PW\PW,\si_1\si_4\si_3\si_2}$.
Note that in the notation we have suppressed the momentum arguments
which, however, can be trivially restored, because the permutation of
momenta $k_i$ is the same as for the polarizations $\si_i$ in each
amplitude.

\subsection{Matrix element for real-gluon emission $\PH\to4f\Pg$}
\label{se:calcrcs}

The real-gluonic corrections are induced by the process
\beq
\PH(p)
 \;\longrightarrow\;
f_1(k_1,\si_1) + \bar f_2(k_2,\si_2) + f_3(k_3,\si_3) + \bar
f_4(k_4,\si_4) + \Pg(k,\lambda),  
\label{eq:h4fa}
\eeq
where the momenta and helicities of the external particles are
indicated in parentheses.

The matrix elements for this process can be constructed from the
matrix elements for the photon radiation process
$\M^{\si_a\si_b\si_c\si_d\la}_{\ga}(Q_a,Q_b,Q_c,Q_d,k_a,k_b,k_c,k_d,k)$,
which have been explicitly given in \citere{Bredenstein:2006rh}.
Here, $Q_{a,b,c,d}$ denote the electric charges of the fermions.  The
generic amplitudes read
\beqar\label{eq:MEH4fa}
\lefteqn{\A^{VV,\si_a\si_b\si_c\si_d\la,h}_{\Pg,c_ac_bc_cc_d}(k_a,k_b,k_c,k_d,k) =} 
\qquad\nl
&&
\frac{\gs}{e}\Biggl\{
\frac{1}{2}\la^h_{c_ac_b}\de_{c_cc_d}\de_{f_aq}
\M^{VV,\si_a\si_b\si_c\si_d\la}_{\gamma}(1,1,0,0,k_a,k_b,k_c,k_d,k)
\nl&& \quad {}
+\frac{1}{2}\la^h_{c_cc_d}\de_{c_ac_b}\de_{f_cq}
\M^{VV,\si_a\si_b\si_c\si_d\la}_{\gamma}(0,0,1,1,k_a,k_b,k_c,k_d,k)
\Biggr\},
\eeqar
where 
$\gs$ is the strong coupling constant, and $V=\PZ,\PW$ for
$\PZ$-mediated and $\PW$-mediated decays, respectively. The symbols
$\de_{f_iq}$ are equal to one if $f_i$ is a quark and zero otherwise.

From the generic matrix element
$\A_{\Pg,c_ac_bc_cc_d}^{VV,\si_a\si_b\si_c\si_d\la,h}(k_a,k_b,k_c,k_d,k)$
the matrix elements for the specific processes can be constructed as
follows.  As above, we denote different fermions ($f\ne F,F'$) by $f$ and
$F$, and their weak-isospin partners by $f'$ and $F'$, respectively:
\begin{itemize}
\item $\PH\to f\bar f F\bar F\Pg$:
\beqar
\A_{\Pg,c_1c_2c_3c_4}^{\si_1\si_2\si_3\si_4\la,h}(k_1,k_2,k_3,k_4,k)&=&
\A_{\Pg,c_1c_2c_3c_4}^{\ZZ,\si_1\si_2\si_3\si_4\la,h}(k_1,k_2,k_3,k_4,k),
\label{eq:zz}
\eeqar
\item $\PH\to f\bar f' F\bar F'\Pg$:
\beqar
\A_{\Pg,c_1c_2c_3c_4}^{\si_1\si_2\si_3\si_4\la,h}(k_1,k_2,k_3,k_4,k)
&=&\A_{\Pg,c_1c_2c_3c_4}^{\WW,\si_1\si_2\si_3\si_4\la,h}(k_1,k_2,k_3,k_4,k),
\label{eq:ww}
\eeqar
\item $\PH\to q\bar q q\bar q\Pg$:
\beqar
\A_{\Pg,c_1c_2c_3c_4}^{\si_1\si_2\si_3\si_4\la,h}(k_1,k_2,k_3,k_4,k)&=&
\A_{\Pg,c_1c_2c_3c_4}^{\ZZ,\si_1\si_2\si_3\si_4\la,h}(k_1,k_2,k_3,k_4,k)
\nl&&{}
-\A_{\Pg,c_1c_4c_3c_2}^{\ZZ,\si_1\si_4\si_3\si_2\la,h}(k_1,k_4,k_3,k_2,k),
\eeqar
\item $\PH\to q\bar q q'\bar q'\Pg$:
\beqar
\A_{\Pg,c_1c_2c_3c_4}^{\si_1\si_2\si_3\si_4\la,h}(k_1,k_2,k_3,k_4,k)
&=&\A_{\Pg,c_1c_2c_3c_4}^{\ZZ,\si_1\si_2\si_3\si_4\la,h}(k_1,k_2,k_3,k_4,k)
\nl&&{}
-\A_{\Pg,c_1c_4c_3c_2}^{\WW,\si_1\si_4\si_3\si_2\la,h}(k_1,k_4,k_3,k_2,k).
\label{eq:mixed}
\eeqar
\end{itemize}
The relative signs between contributions of the basic subamplitudes to
the full matrix elements account for the sign changes resulting from
interchanging external fermion lines.

Squaring the amplitudes and summing over the colour degrees of
freedom, we have
\begin{itemize}
\item $\PH\to f\bar f F\bar F\Pg$:
\beqar
\lefteqn{
\sum_{\{c_i\},h}
\left|\A_{\Pg,c_1c_2c_3c_4}^{\si_1\si_2\si_3\si_4\la,h}(k_1,k_2,k_3,k_4,k)\right|^2 =
\frac{4}{3}N^{\mathrm{c}}_{f}N^{\mathrm{c}}_{F}
\frac{\als}{\al}
}\qquad&&\hspace{12em}
\nl&&{}\times
\Biggl[\de_{fq}
\left|\M_{\ga}^{\ZZ,\si_1\si_2\si_3\si_4\la}(1,1,0,0,k_1,k_2,k_3,k_4,k)\right|^2 
\nl&& \quad {}+\de_{Fq}
\left|\M_{\ga}^{\ZZ,\si_1\si_2\si_3\si_4\la}(0,0,1,1,k_1,k_2,k_3,k_4,k)\right|^2 
\Biggr],
\hspace*{12em}
\eeqar
\item $\PH\to f\bar f' F\bar F'\Pg$:
\beqar
\lefteqn{
\sum_{\{c_i\},h}
\left|\A_{\Pg,c_1c_2c_3c_4}^{\si_1\si_2\si_3\si_4\la,h}(k_1,k_2,k_3,k_4,k)\right|^2 =
\frac{4}{3}N^{\mathrm{c}}_{f}N^{\mathrm{c}}_{F}
\frac{\als}{\al}
}\qquad&&\hspace{12em}
\nl&&{}\times
\Biggl[\de_{fq}
\left|\M_{\ga}^{\WW,\si_1\si_2\si_3\si_4\la}(1,1,0,0,k_1,k_2,k_3,k_4,k)\right|^2 
\nl&& \quad {}+\de_{Fq}
\left|\M_{\ga}^{\WW,\si_1\si_2\si_3\si_4\la}(0,0,1,1,k_1,k_2,k_3,k_4,k)\right|^2 
\Biggr],
\hspace*{12em}
\eeqar
\item $\PH\to q\bar q q\bar q\Pg$:
\beqar
\lefteqn{
\sum_{\{c_i\},h}
\left|\A_{\Pg,c_1c_2c_3c_4}^{\si_1\si_2\si_3\si_4\la,h}(k_1,k_2,k_3,k_4,k)\right|^2 =
\frac{4}{3}(N^{\mathrm{c}}_{q})^2
\frac{\als}{\al}
}\qquad\nl&&{}
\times\biggl[
\left|\M_{\ga}^{\ZZ,\si_1\si_2\si_3\si_4\la}(1,1,0,0,k_1,k_2,k_3,k_4,k)\right|^2 
\nl&& \quad {}
+\left|\M_{\ga}^{\ZZ,\si_1\si_2\si_3\si_4\la}(0,0,1,1,k_1,k_2,k_3,k_4,k)\right|^2 
\nl&& \quad {}
+\left|\M_{\ga}^{\ZZ,\si_1\si_4\si_3\si_2\la}(1,1,0,0,k_1,k_4,k_3,k_2,k)\right|^2 
\nl&& \quad {}
+\left|\M_{\ga}^{\ZZ,\si_1\si_4\si_3\si_2\la}(0,0,1,1,k_1,k_4,k_3,k_2,k)\right|^2 
\biggr]
\nl&&{}
-\frac{8}{3}N^{\mathrm{c}}_{q}
\frac{\als}{\al}\Re\biggl[
\Bigl(\M_{\ga}^{\ZZ,\si_1\si_2\si_3\si_4\la}(1,1,0,0,k_1,k_2,k_3,k_4,k)
\nl&&{}\qquad
+\M_{\ga}^{\ZZ,\si_1\si_2\si_3\si_4\la}(0,0,1,1,k_1,k_2,k_3,k_4,k)\Bigr)^*
\nl&&{}\quad\times
\Bigl(\M_{\ga}^{\ZZ,\si_1\si_4\si_3\si_2\la}(1,1,0,0,k_1,k_4,k_3,k_2,k)
\nl&&{}\qquad
+\M_{\ga}^{\ZZ,\si_1\si_4\si_3\si_2\la}(0,0,1,1,k_1,k_4,k_3,k_2,k)\Bigr)
\Biggr],
\hspace*{12em}
\label{eq:M2gzzsym}
\eeqar
\item $\PH\to q\bar q q'\bar q'\Pg$:
\beqar
\lefteqn{
\sum_{\{c_i\},h}
\left|\A_{\Pg,c_1c_2c_3c_4}^{\si_1\si_2\si_3\si_4\la,h}(k_1,k_2,k_3,k_4,k)\right|^2 =
\frac{4}{3}(N^{\mathrm{c}}_{q})^2
\frac{\als}{\al}}
\qquad\nl&&{}
\times\biggl[
\left|\M_{\ga}^{\ZZ,\si_1\si_2\si_3\si_4\la}(1,1,0,0,k_1,k_2,k_3,k_4,k)\right|^2 
\nl&& \quad {}
+\left|\M_{\ga}^{\ZZ,\si_1\si_2\si_3\si_4\la}(0,0,1,1,k_1,k_2,k_3,k_4,k)\right|^2 
\nl&& \quad {}
+\left|\M_{\ga}^{\WW,\si_1\si_4\si_3\si_2\la}(1,1,0,0,k_1,k_4,k_3,k_2,k)\right|^2 
\nl&& \quad {}
+\left|\M_{\ga}^{\WW,\si_1\si_4\si_3\si_2\la}(0,0,1,1,k_1,k_4,k_3,k_2,k)\right|^2 
\biggr]
\nl&&{}
-\frac{8}{3}N^{\mathrm{c}}_{q}
\frac{\als}{\al}\Re\biggl[
\Bigl(\M_{\ga}^{\ZZ,\si_1\si_2\si_3\si_4\la}(1,1,0,0,k_1,k_2,k_3,k_4,k)
\nl&&{}\qquad
+\M_{\ga}^{\ZZ,\si_1\si_2\si_3\si_4\la}(0,0,1,1,k_1,k_2,k_3,k_4,k)\Bigr)^*
\nl&&{}\quad\times
\Bigl(\M_{\ga}^{\WW,\si_1\si_4\si_3\si_2\la}(1,1,0,0,k_1,k_4,k_3,k_2,k)
\nl&&{}\qquad
+\M_{\ga}^{\WW,\si_1\si_4\si_3\si_2\la}(0,0,1,1,k_1,k_4,k_3,k_2,k)\Bigr)
\biggr].
\hspace*{12em}
\label{eq:M2gmixed}
\eeqar
\end{itemize}

The contribution $\Ga_\Pg$ of the radiative decay to the total
decay width is given by
\beq
\Ga_\Pg = \frac{1}{2\MH} \int \rd\Phi_\Pg \,
 \sum_{\{c_i\},h}
\, \sum_{\{\si_i\},\lambda=\pm 1} \,
|\A^{\si_1\si_2\si_3\si_4\la,h}_{\Pg,c_1c_2c_3c_4}|^2,
\label{eq:hbcs}
\eeq
where the phase-space integral is defined by
\beq
\int \rd\Phi_\Pg =
\int\frac{\rd^3 {\bf k}}{(2\pi)^3 2k^0} \,
\left( \prod_{i=1}^4 \int\frac{\rd^3 {\bf k}_i}{(2\pi)^3 2k_i^0} \right)\,
(2\pi)^4 \delta\Biggl(p-k-\sum_{j=1}^4 k_j\Biggr).
\label{eq:dPSg}
\eeq

\section{Numerical results}
\label{se:numerics}

\subsection{Setup and input}
\label{se:numsetup}

\begin{sloppypar} 
We use the $G_\mu$ scheme, i.e.\ we define the electromagnetic
coupling by $\alpha_{G_\mu}={\sqrt{2}G_\mu \MW^2(1-\MW^2/\MZ^2)}/{\pi}$.  Our
lowest-order results include the ${\cal O}(\alpha)${}-corrected width
of the gauge bosons. In the QCD corrections we uniformly take a fixed
value for $\als=\als(\MZ)=0.1187$ everywhere, because the only
numerically relevant part (see below) of the QCD correction is the one
connected with the hadronic decay of a W or a Z~boson, where the scale
is fixed by the intermediate gauge-boson decay.  More details about
the setup and all input parameters are provided in
\citere{Bredenstein:2006rh}.  
\end{sloppypar}

In our approach the final states involve either four fermions (from
lowest order and virtual corrections), four fermions and a photon
(from real-photonic corrections), four fermions and a gluon (from
real-gluonic corrections), or four fermions and one or more photons
collinear to an outgoing lepton (from the structure functions
describing multi-photon final-state radiation).  In particular there
are no events containing both photons and gluons.  Moreover, we only
consider semileptonic final states in the distributions.  For these
distributions, a photon and gluon recombination is performed as
follows.  In events with a real photon, as in
\citere{Bredenstein:2006rh} the photon is recombined with the (in this
sense) nearest charged fermion if the invariant mass of the
photon--fermion pair is below $5\GeV$.  This, in particular, implies
that all photons collinear to a lepton are recombined with the
corresponding lepton if the recombination is switched on (as always
done in the results of this paper), i.e.\ the higher-order effects
from photonic final-state radiation described in Section~4.3 of
\citere{Bredenstein:2006rh} fully cancel out in this case.  In the
case of real-gluon radiation we force a 2-jet event.  This is achieved
by always recombining the two partons of the $qq\Pg$ system that yield
the smallest invariant mass.  Invariant masses and angles are then
defined by the 4-momenta of the recombined pair and the remaining
partons.

We always sum over the quarks of the first two generations,
$q=\Pu,\Pd,\Pc,\Ps$, and over the three neutrinos in the final states
and we consider the final states $\mathrm{ee}qq$, $\nu\nu qq$,
$\mathrm{e}\nu qq$, and $qqqq$.  Since we consistently neglect the
masses of external fermions and average over polarizations, we can
express the partial widths of these final states as
\beqar
\Gamma_{\PH\to\Pe\Pe qq} &=& 2\Gamma_{\PH\to\Pem\Pep\Pu\Pubar} 
                       + 2\Gamma_{\PH\to\Pem\Pep\Pd\Pdbar},
\nn\\
\Gamma_{\PH\to\nu\nu qq} &=& 6\Gamma_{\PH\to\Pne\Pnebar\Pu\Pubar} 
                       + 6\Gamma_{\PH\to\Pne\Pnebar\Pd\Pdbar},
\nn\\
\Gamma_{\PH\to\Pe\nu qq} &=& 4\Gamma_{\PH\to\Pem\Pnebar\Pu\Pdbar},
\nn\\
\Gamma_{\PH\to qqqq} &=& \Gamma_{\PH\to\Pu\Pubar\Pc\Pcbar} 
                        + \Gamma_{\PH\to\Pd\Pdbar\Ps\Psbar}
                        + 2\Gamma_{\PH\to\Pu\Pubar\Ps\Psbar} 
                        + 2\Gamma_{\PH\to\Pu\Pdbar\Ps\Pcbar}
\nn\\
&&{}                    + 2\Gamma_{\PH\to\Pu\Pdbar\Pd\Pubar} 
                        + 2\Gamma_{\PH\to\Pu\Pubar\Pu\Pubar}
                        + 2\Gamma_{\PH\to\Pd\Pdbar\Pd\Pdbar}.
\eeqar
Note that $\Gamma_{\PH\to\Pe\nu qq}$ includes both electrons and
positrons in the final state. The partial widths with muons in the
final state can be classified in the same way and are
equal to those with the muons replaced by electrons,
because no dependence on the final-state fermion masses remains
for these inclusive quantities.

The results for the partial decay widths in the plots are calculated
using $10^7$ Monte Carlo events, while all other results (decay widths
in the table and distribution plots) are obtained with $5\times 10^7$
events.  In the presented results, soft and collinear divergences are
treated with the dipole-subtraction method and have been checked by
applying the phase-space slicing method.  For the latter method more
Monte Carlo events are needed for an accuracy at the per-mille level,
because the energy and angular cuts in this method have to be chosen
small enough rendering the real corrections and the analytically
integrated soft and collinear singular contribution (which compensate
each other) very large.  In both methods it is possible to evaluate
the virtual corrections (rendered finite by adding the soft and
collinear singularities from the real corrections) less often than the
lowest-order matrix elements, because the virtual corrections and also
their statistical error are smaller. We evaluate the EW
virtual corrections only every 100th time and the virtual QCD
corrections only every 20th time. This procedure reduces the run-time
of the program while maintaining the size of the overall statistical
error.

\subsection{Results for partial decay widths}

In \refta{tab:width} we show the partial decay widths of the Higgs
boson for semileptonic and hadronic final states for different values
of the Higgs-boson mass.
\begin{table}
\centerline{
\begin{tabular}{|c|c|c|c|c|c|c|c|}
\hline
& $\MH[\GeV]$ & \multicolumn{2}{c|}{$140$} 
 & \multicolumn{2}{c|}{$170$} 
 & \multicolumn{2}{c|}{$200$} 
 \\
\hline
\hline
& $\GW[\GeV]$ & \multicolumn{2}{c|}{$2.09052...$} 
 & \multicolumn{2}{c|}{$2.09054...$} 
 & \multicolumn{2}{c|}{$2.09055...$} 
 \\
\hline
& $\GZ[\GeV]$ & \multicolumn{2}{c|}{$2.50278...$} 
 & \multicolumn{2}{c|}{$2.50287...$} 
 & \multicolumn{2}{c|}{$2.50292...$} 
 \\
\hline
\hline
$\PH\;\to$& & $\Gamma [\MeV]$ & $\delta[\%]$ & 
$\Gamma [\MeV]$ & $\delta[\%]$ & $\Gamma [\MeV]$ & $\delta[\%]$\\
\hline
                         $\mathrm{ee}qq$                         
 & corrected
 &    0.020467(6)
 &    5.1
 &    0.32723(9)
 &    5.7
 &   13.332(2)
 &    7.6
  \\
 & EW
 &    0.019731(5)
 &    1.3
 &    0.31558(7)
 &    2.0       
 &   12.863(1)
 &    3.8
  \\
 & QCD
 &    0.020217(5)
 &    3.8
 &    0.32115(7)
 &    3.8
 &   12.858(1)
 &    3.8
  \\
 & LO
 &    0.019481(4)
 &
 &    0.30950(5)
 &
 &   12.389(1)
 &
  \\\hline
                         $\nu\nu qq$                             
 & corrected
 &    0.12221(4)
 &    5.9
 &    1.9559(6)
 &    6.7
 &   79.69(1)
 &    8.5
  \\
 & EW
 &    0.11784(3)
 &    2.1
 &    1.8873(4)
 &    2.9
 &   76.91(1)
 &    4.8
  \\
 & QCD
 &    0.11982(3)
 &    3.8
 &    1.9025(5)
 &    3.7
 &   76.20(1)
 &    3.8
  \\
 & LO
 &    0.11545(3)
 &
 &    1.8339(4)
 &
 &   73.423(8)
 &
  \\\hline
                         $\mathrm{e}\nu qq$                      
 & corrected
 &    0.5977(3)
 &    7.4
 &   53.55(2)
 &    9.9
 &  155.37(4)
 &    8.7
  \\
 & EW
 &    0.5767(2)
 &    3.6
 &   51.71(1)
 &    6.1
 &  149.96(3)
 &    4.9
  \\
 & QCD
 &    0.5775(2)
 &    3.8
 &   50.57(1)
 &    3.8    
 &  148.32(3)
 &    3.8    
  \\
 & LO
 &    0.5564(2)
 &
 &   48.724(9)
 &
 &  142.91(2)
 &
  \\\hline
                         $qqqq$                                  
 & corrected
 &    2.0113(8)
 &   10.8
 &  168.73(5)
 &   13.6
 &  590.3(1)
 &   12.1
  \\
 & EW
 &    1.8752(4)
 &    3.3
 &  157.50(2)
 &    6.0    
 &  550.47(7)
 &    4.6
  \\
 & QCD
 &    1.9511(7)
 &    7.5
 &  159.83(4)
 &    7.6
 &  566.2(1)
 &    7.6
  \\
 & LO
 &    1.8150(4)
 &
 &  148.59(2)
 &
 &  526.39(5)
 &
  \\\hline
\end{tabular} }
\caption{Partial decay widths $\Ga_{\PH\to4f}$ in lowest order (LO),
  including \Oa{} and ${\cal O}(\GF^2\MH^4)$ EW corrections, \Oas{}
  QCD corrections, and the sum of EW and QCD corrections (corrected)
  and corresponding relative corrections $\de$ for semileptonic and
  hadronic decay channels and different Higgs-boson masses.} 
\label{tab:width}
\end{table}
We list the lowest-order (LO) predictions and the predictions
including the complete EW \Oa{} plus ${\cal O}(\GF^2\MH^4)$
corrections and the \Oas{} QCD corrections. In addition we give the
predictions including only the EW corrections and only the QCD
corrections. In all cases we provide also the relative corrections
$\de=\Ga/\Ga_0-1$ in per cent.  The statistical errors of the
phase-space integration are given in parentheses.  The size of the EW
corrections is very similar to the size of the corresponding
corrections for leptonic final states discussed in
\citere{Bredenstein:2006rh}.  Since the QCD corrections mainly arise
from vertex corrections and since we consider the integrated partial
widths, the QCD contribution roughly amounts to $\als/\pi$ for
semileptonic final states and $2\als/\pi$ for the hadronic final
state.  The sum of EW and QCD corrections thus rises to $5$--$14\%$.

\begin{figure}
\setlength{\unitlength}{1cm}
\centerline{
\begin{picture}(7.7,8)
\put(-1.7,-14.5){\includegraphics{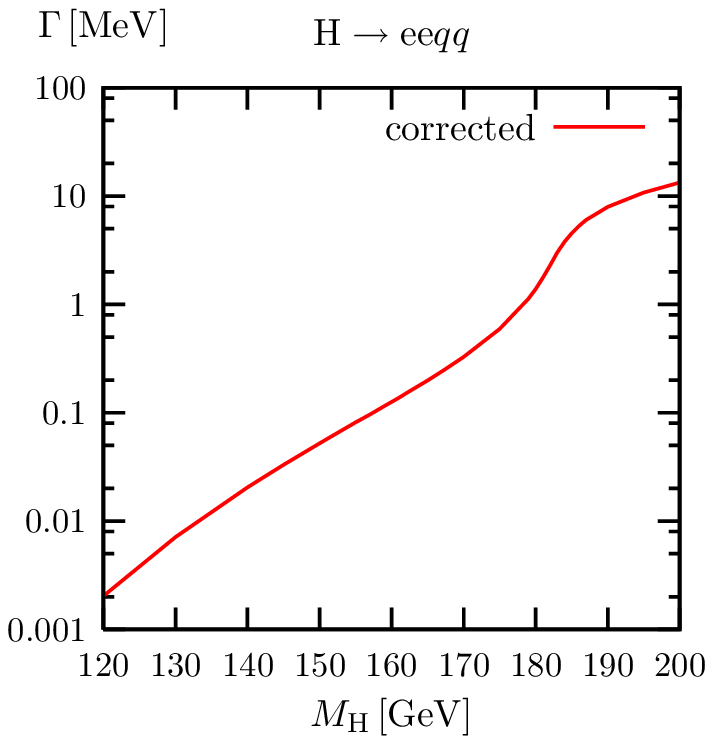}}
\end{picture}
\begin{picture}(7.5,8)
\put(-1.7,-14.5){\includegraphics{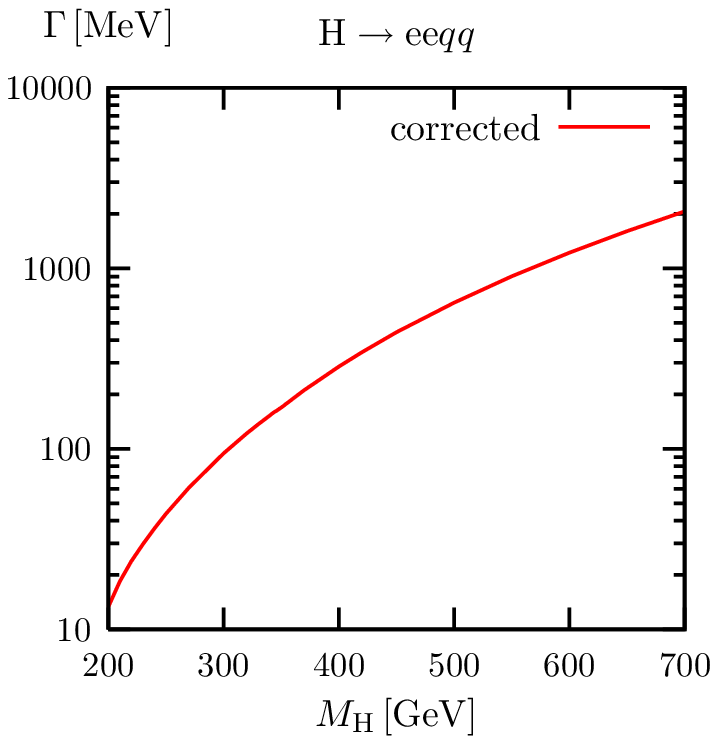}}
\end{picture} }
\centerline{
\begin{picture}(7.7,8)
\put(-1.7,-14.5){\includegraphics{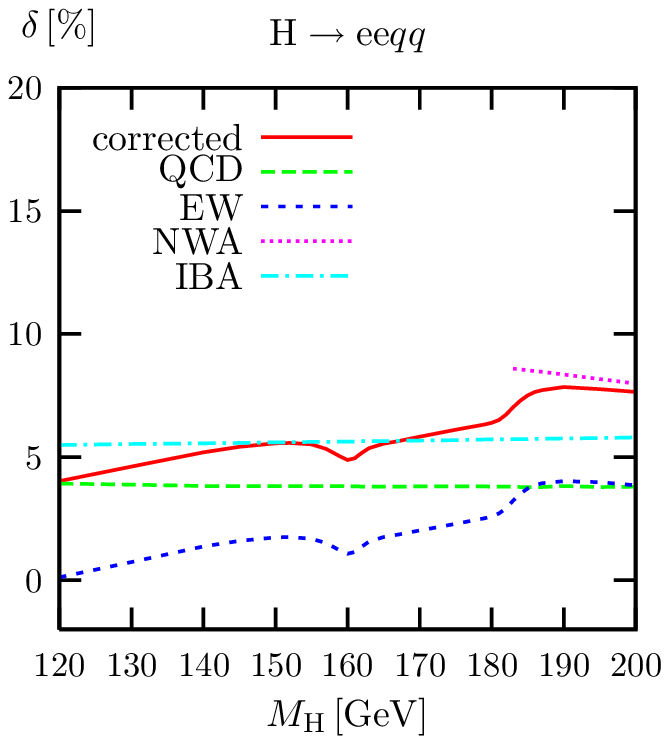}}
\end{picture}
\begin{picture}(7.5,8)
\put(-1.7,-14.5){\includegraphics{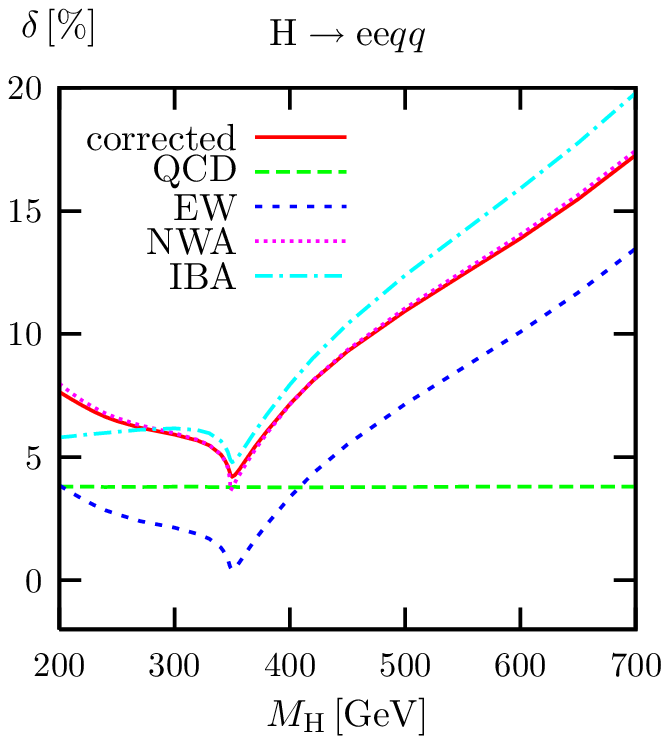}}
\end{picture} }
\caption{Partial decay width for $\PH\to\Pe\Pe qq$ 
  as a function of the Higgs-boson mass. The upper plots show the
  absolute prediction including QCD and EW corrections, and the lower
  plots show the relative size of the QCD and EW corrections separately,
  their sum (corrected) and the predictions of the NWA and the IBA.}
\label{fig:eeqq}
\end{figure}

\begin{figure}
\setlength{\unitlength}{1cm}
\centerline{
\begin{picture}(7.7,8)
\put(-1.7,-14.5){\includegraphics{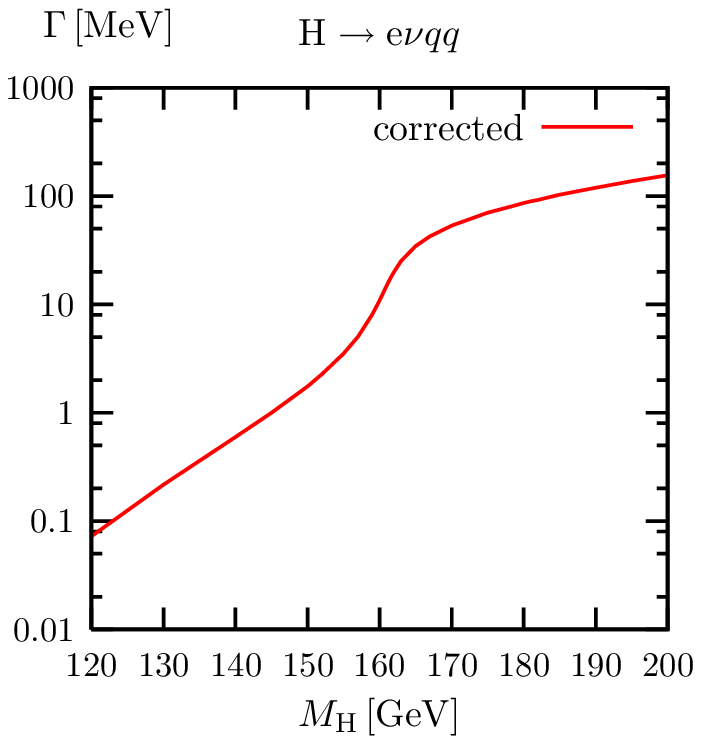}}
\end{picture}
\begin{picture}(7.5,8)
\put(-1.7,-14.5){\includegraphics{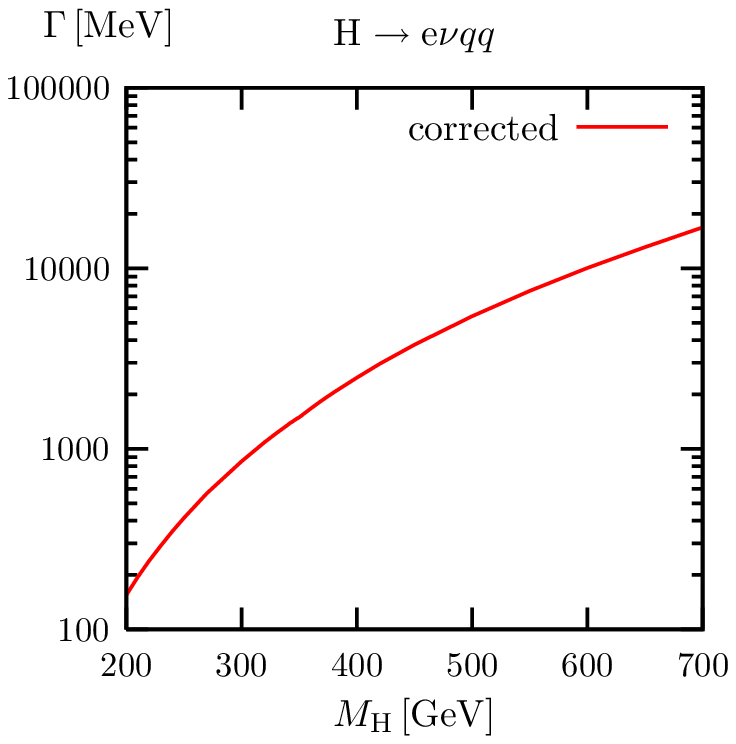}}
\end{picture} }
\centerline{
\begin{picture}(7.7,8)
\put(-1.7,-14.5){\includegraphics{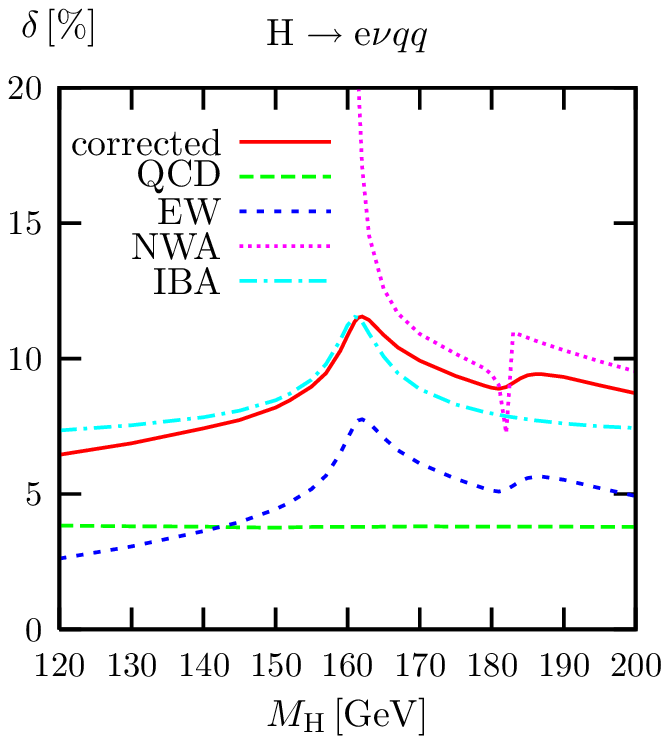}}
\end{picture}
\begin{picture}(7.5,8)
\put(-1.7,-14.5){\includegraphics{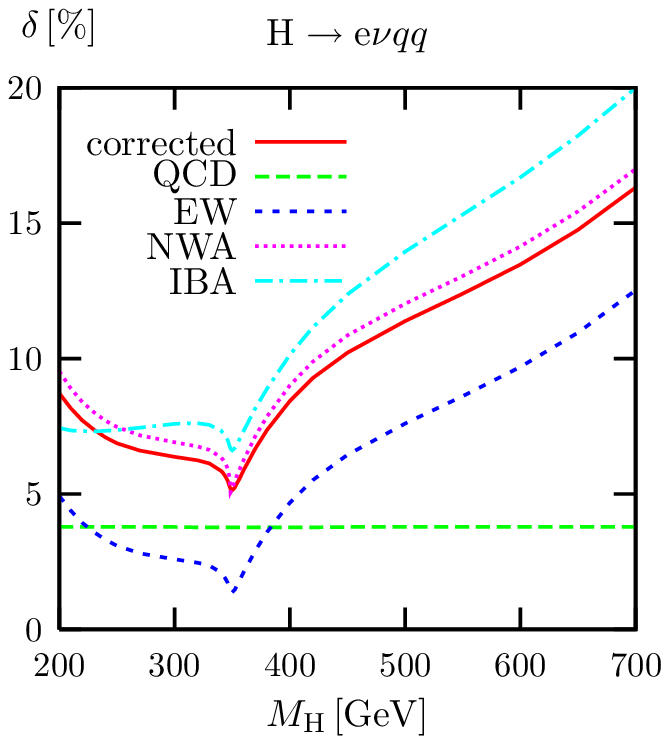}}
\end{picture} }
\caption{Partial decay width for $\PH\to\Pe\nu qq$ as a function of
  the Higgs-boson mass. The individual curves are defined as in
  \reffi{fig:eeqq}.} 
\label{fig:enuqq}
\end{figure}

\begin{figure}
\setlength{\unitlength}{1cm}
\centerline{
\begin{picture}(7.7,8)
\put(-1.7,-14.5){\includegraphics{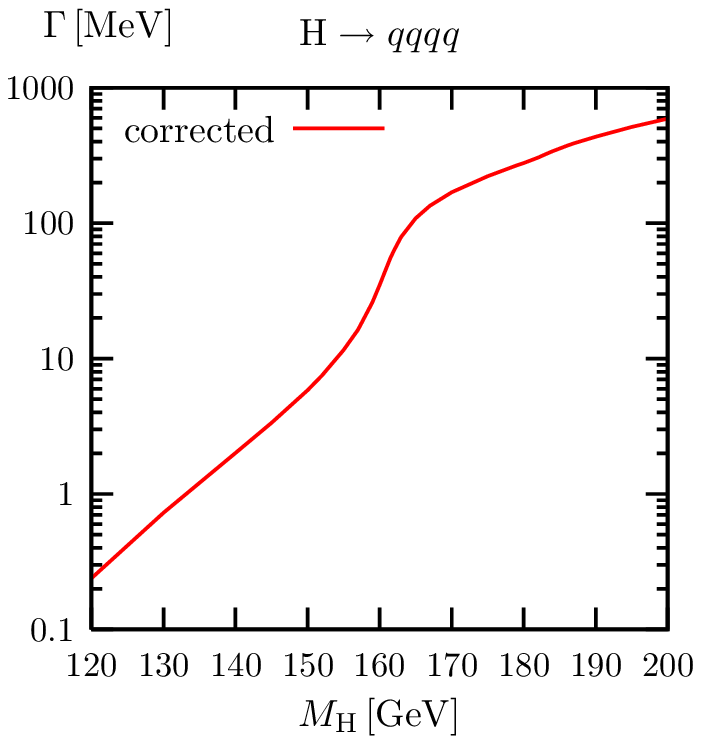}}
\end{picture}
\begin{picture}(7.5,8)
\put(-1.7,-14.5){\includegraphics{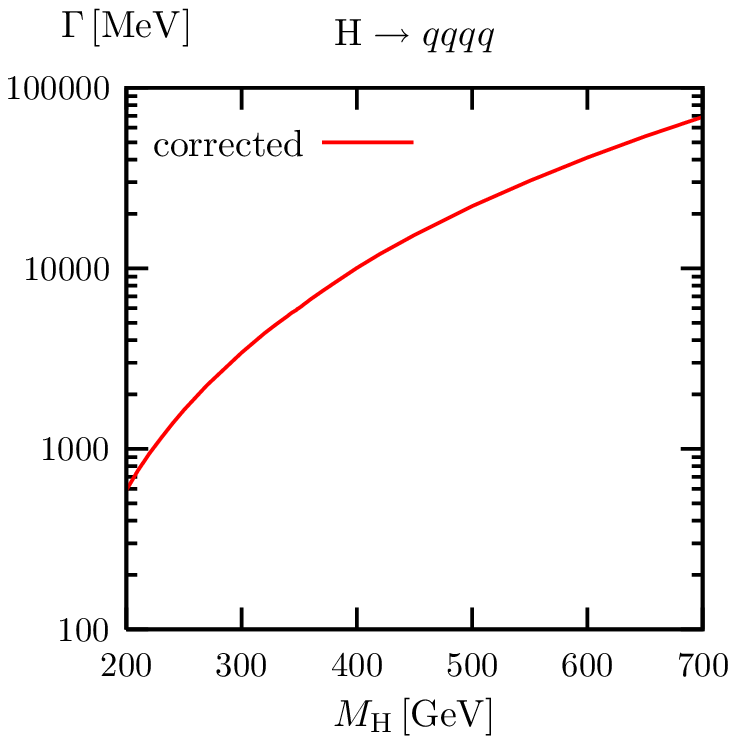}}
\end{picture} }
\centerline{
\begin{picture}(7.7,8)
\put(-1.7,-14.5){\includegraphics{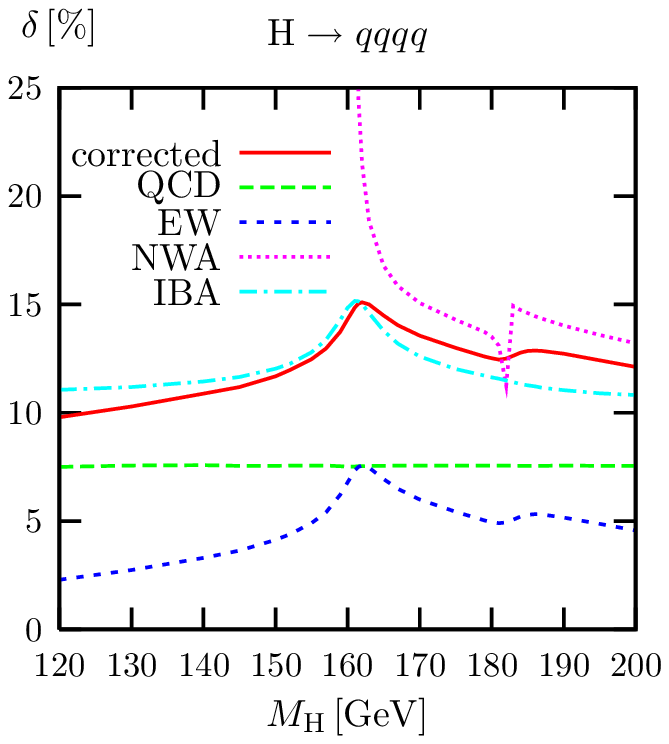}}
\end{picture}
\begin{picture}(7.5,8)
\put(-1.7,-14.5){\includegraphics{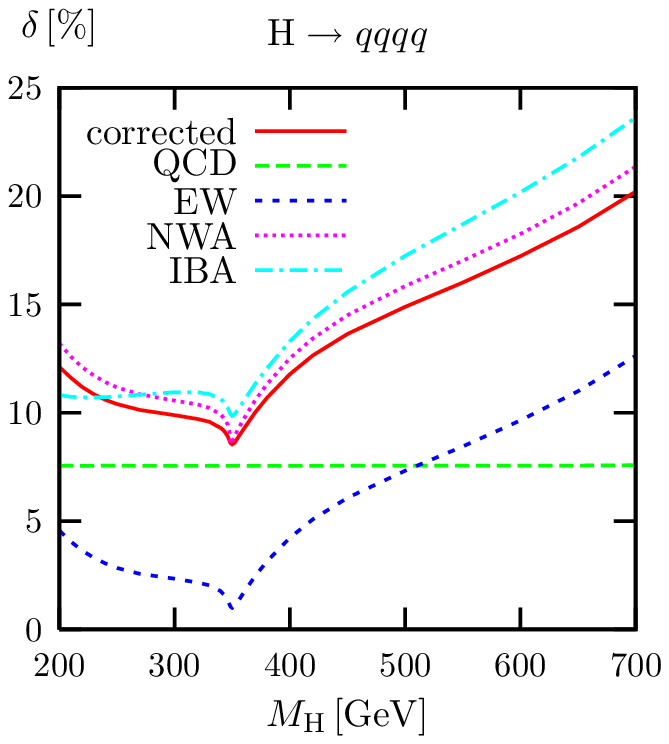}}
\end{picture} }
\caption{Partial decay width for $\PH\to qqqq$ 
  as a function of the Higgs-boson mass.  The individual curves are
  defined as in \reffi{fig:eeqq}.}
\label{fig:qqqq}
\end{figure}

In \reffis{fig:eeqq}, \ref{fig:enuqq}, and \ref{fig:qqqq} we show the
partial decay widths as a function of the Higgs-boson mass for
$\PH\to\Pe\Pe qq$, $\PH\to\Pe\nu qq$, and $\PH\to qqqq$, respectively.
The upper plots show the predictions including both QCD and EW
corrections. The lower plots depict the corrections relative to the
lowest order. Besides the EW+QCD corrections, these plots include the
EW and QCD corrections separately, the narrow-width approximation
(NWA) and the improved Born approximation (IBA) as defined in
Eqs.~(7.5)--(7.7) and Eqs.~(6.1)--(6.7), respectively, of
\citere{Bredenstein:2006rh}. We recall that the IBA for the partial
decay widths includes leading effects such as corrections that are
enhanced by factors $\GF\Mt^2$ or $\GF\MH^2$, the Coulomb singularity
for W~pairs near their on-shell threshold, and the QCD correction to
hadronically decaying gauge bosons.  Apart from these effects, the IBA
contains only one fitted constant for the WW- and ZZ-mediated channels
each.  Both in the WW-induced channel and in the ZZ-induced channel
the EW corrections are very similar to the corresponding corrections
for leptonic final states \cite{Bredenstein:2006rh}.  For moderate
Higgs-boson mass, they are positive and below $\sim4\%$ for
decays via Z pairs.  For the W-mediated decays the Coulomb singularity
yields a large effect near the WW threshold and the EW corrections are
in the range between 2\% and 8\% for moderate Higgs-boson mass. For
all decays the EW corrections reach about $13\%$ near $\MH=700\GeV$.
The thresholds for the on-shell decay of the Higgs boson into W
bosons, Z bosons, and top quarks are manifest in the shape of the
corrections.  The QCD corrections amount to roughly $\als/\pi\approx
3.8\%$ for semileptonic and $2\als/\pi\approx 7.6\%$ for hadronic
final states and are practically independent of the Higgs-boson mass.
For $\PH\to\Pe\Pe qq$ the agreement between the full result and the
NWA is (accidentally) at the per-mille level sufficiently above the ZZ
threshold. For $\PH\to\Pe\nu qq$ and $\PH\to qqqq$ the NWA agrees with
the full results within 1--2\% above threshold. The IBA describes the
full corrections within 2--3\% for $\MH\lsim400\GeV$ for all final
states.

\begin{figure}
\setlength{\unitlength}{1cm}
\centerline{
\begin{picture}(7.7,8)
\put(-1.7,-14.5){\includegraphics{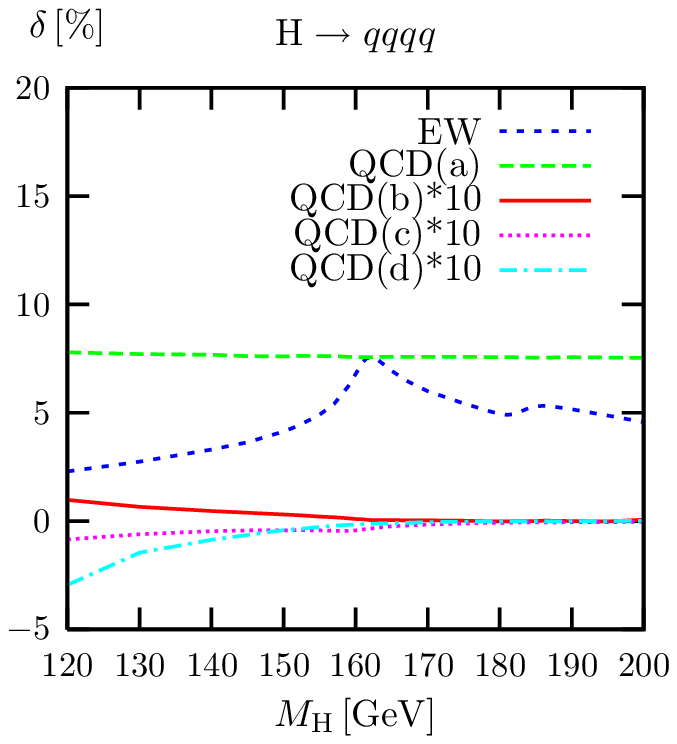}}
\end{picture}
\begin{picture}(7.5,8)
\put(-1.7,-14.5){\includegraphics{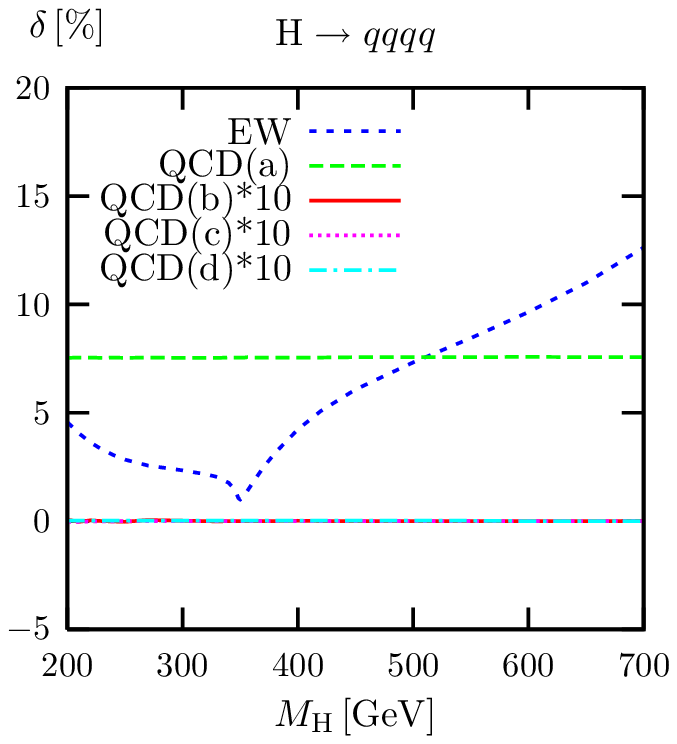}}
\end{picture} }
\caption{Comparison of the different QCD contributions defined in 
  \refse{se:class} and the EW contribution to the corrections to the 
  partial decay width for $\PH\to qqqq$ as a function of the Higgs-boson mass.}
\label{fig:abcd}
\end{figure}

In \refse{se:class} we defined a classification of QCD corrections for
the four-quark final states.  While only category (a), \ie QCD
corrections to gauge-boson decays, exists for the final states $\PH\to
q\bar q Q\bar Q$ and $\PH\to q\bar q' Q\bar Q'$ ($q\ne Q,Q'$), all
categories (a)--(d) contribute to $\PH\to q\bar q q\bar q$ and $\PH\to
q\bar q q'\bar q'$.  Figure \ref{fig:abcd} shows the relative EW
corrections and the subcontributions of the different categories of
QCD corrections as a function of the Higgs-boson mass.
The corrections to gauge-boson decays, \ie category (a), make up
practically all of the QCD part.  Note that the contributions (b)--(d)
are multiplied by a factor 10 in the plots.  For $\MH\gsim2\MW$, these
contributions are completely negligible. In this region they are
suppressed by a factor $(\GV/\MV)^2$ with respect to the leading
contributions because they have two propagators less that can become
resonant.  Below the WW threshold this suppression becomes smaller but
at $\MH=120\GeV$ the interference contribution is still rather small
reaching only a few per mille. The largest corrections originate from
intermediate $\Pg^*\Pg^*$ states [category (d)], because these
corrections are proportional to $\alpha^2\als^2$ rather than to
$\alpha^3\als$ as all other QCD corrections.

\subsection{Invariant-mass distributions}

\begin{figure}
\setlength{\unitlength}{1cm}
\centerline{
\begin{picture}(7.7,8)
\put(-1.7,-14.5){\includegraphics{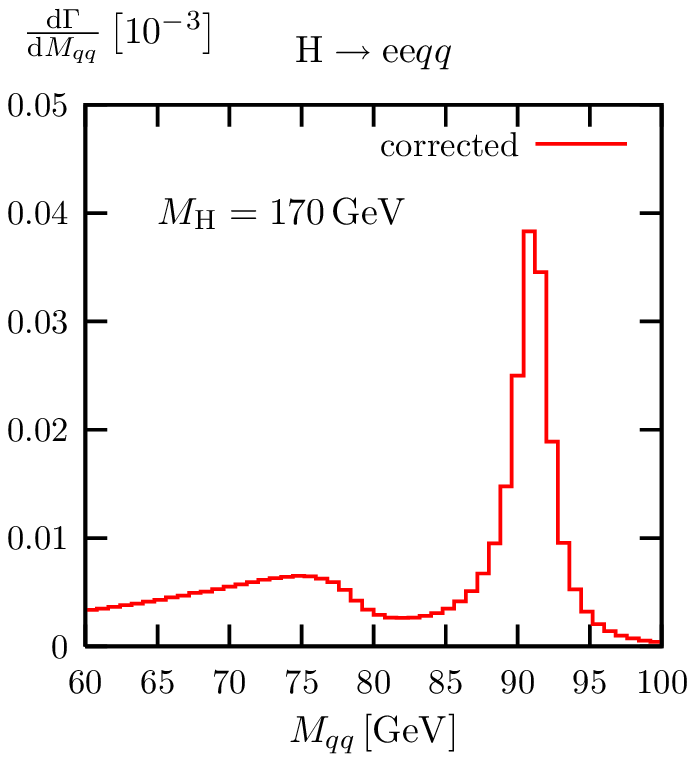}}
\end{picture}
\begin{picture}(7.5,8)
\put(-1.7,-14.5){\includegraphics{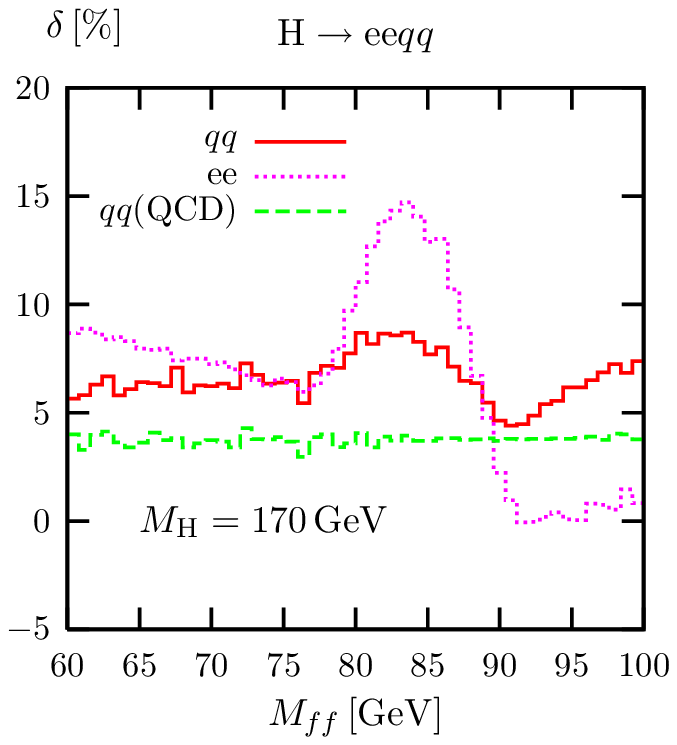}}
\end{picture} }
\centerline{
\begin{picture}(7.7,8)
\put(-1.7,-14.5){\includegraphics{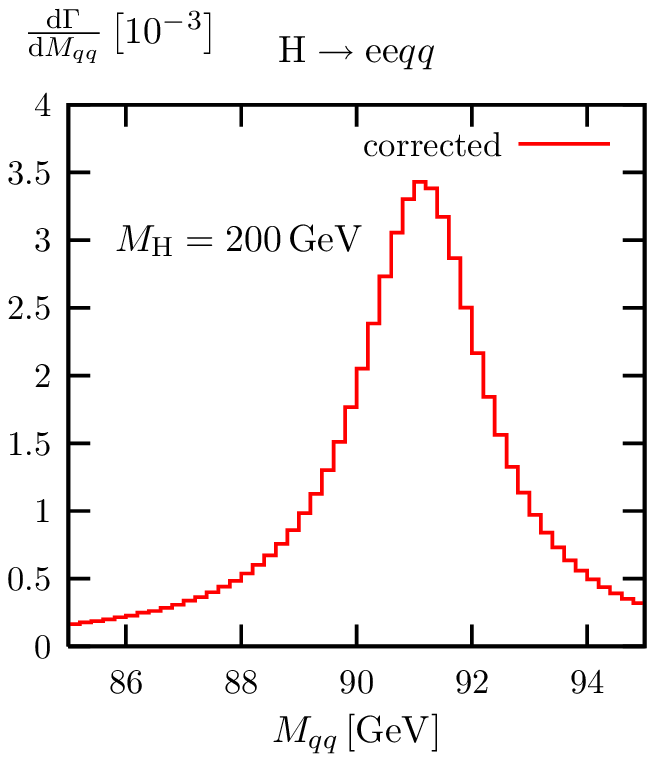}}
\end{picture}
\begin{picture}(7.5,8)
\put(-1.7,-14.5){\includegraphics{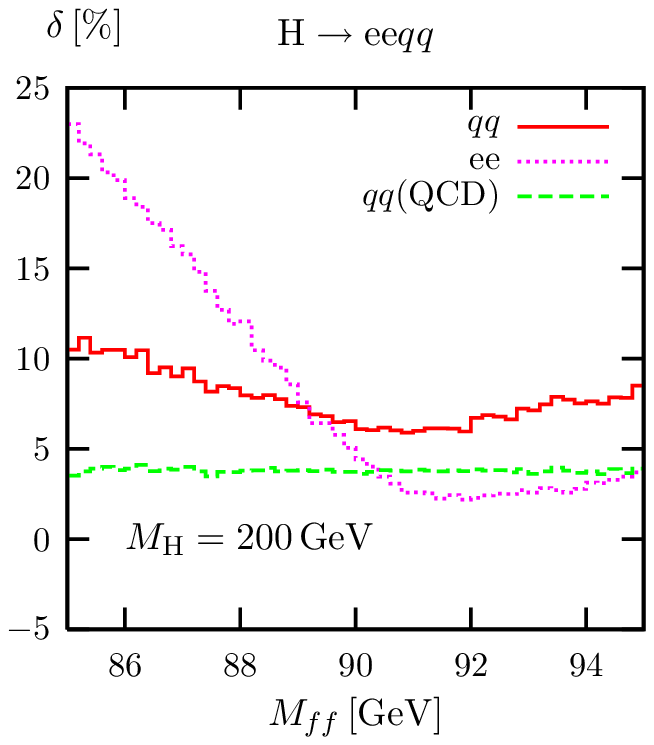}}
\end{picture} }
\caption{Distribution in the invariant mass of the $qq$ pair (l.h.s.) and
  relative EW+QCD corrections to the distributions in the invariant
  mass of the $\mathrm{ee}$ and $qq$ pairs (r.h.s) in the decay
  $\PH\to\mathrm{ee}qq$ for $\MH=170\GeV$ and $\MH=200\GeV$.
  For the distribution in $M_{qq}$ the relative QCD corrections are
  separately shown.}
\label{fig:eeqqinv}
\end{figure}

In order to reconstruct the Higgs-decay events and in order to separate
signal events from possible background events, distributions in the
invariant mass of fermion pairs resulting from a W- or Z-boson decay
should be investigated. On the l.h.s.\ of \reffi{fig:eeqqinv} we show the
invariant-mass distribution of the $qq$ pair 
in the decay $\PH\to\mathrm{ee}qq$ including QCD and EW corrections for
$\MH=170\GeV$ and $\MH=200\GeV$,
i.e.\ for one $\MH$ value below and another above the on-shell threshold
at $2\MZ$ for Z-boson pairs.
Above the threshold for the on-shell decay into a Z-boson pair there
is a just a resonance around the Z-boson mass. Below the threshold
only one Z boson can become resonant while the other Z boson is off
shell. Hence, in addition to the peak around the Z-boson mass, the
$qq$ invariant-mass distribution shows an enhancement for
$M_{qq}<\MH-\MZ\approx 80\GeV$ where the $\Pep\Pem$ 
pair can result from a resonant Z boson.

The complete relative corrections to the distribution in the invariant
mass of the $qq$ pair and also of the $\Pep\Pem$ 
pair are shown on the r.h.s.\ of \reffi{fig:eeqqinv}.  
In addition, the QCD corrections to the $M_{qq}$ distribution
are plotted separately; they are flat and amount to roughly $3.8\%$. 
Note that $M_{qq}$ actually is the total
hadronic invariant mass resulting from the $\PH\to\Pe\Pe qq$ decay, since we
always recombine the $q\bar q(\Pg)$ system to two jets. 
In a detailed experimental analysis a jet
algorithm should be defined. Then, hard gluons can produce a separate
jet and the QCD corrections need not be flat anymore. For such a
study, the jet algorithm could simply be interfaced to 
our Monte Carlo program.
The EW corrections reveal the same structure as discussed in the case
of leptonic decays \cite{Bredenstein:2006rh} shifting the peak
position of the resonance. Close to the resonance and above, the EW
corrections can reach $5{-}10\%$, below the resonance they become larger.
The dominant effect is of photonic origin, leading to more pronounced 
corrections in the case of the leptonic invariant mass $M_{\Pep\Pem}$,
since the electric-charge factors are larger for leptons than for quarks.
The way photons are treated has a strong impact
on the corrections. By performing photon
recombination, as defined in \refse{se:numsetup}, we obtain
collinear-safe observables. Thus, the corrections are of moderate
size. However, for non-collinear-safe observables, i.e.\ if no
photon recombination with leptons were performed, the corrections would
be much larger because of mass-singular corrections proportional to
$\alpha\ln(m_l/\MH)$, as discussed in \citere{Bredenstein:2006rh}.

\begin{figure}
\setlength{\unitlength}{1cm}
\centerline{
\begin{picture}(7.7,8)
\put(-1.7,-14.5){\includegraphics{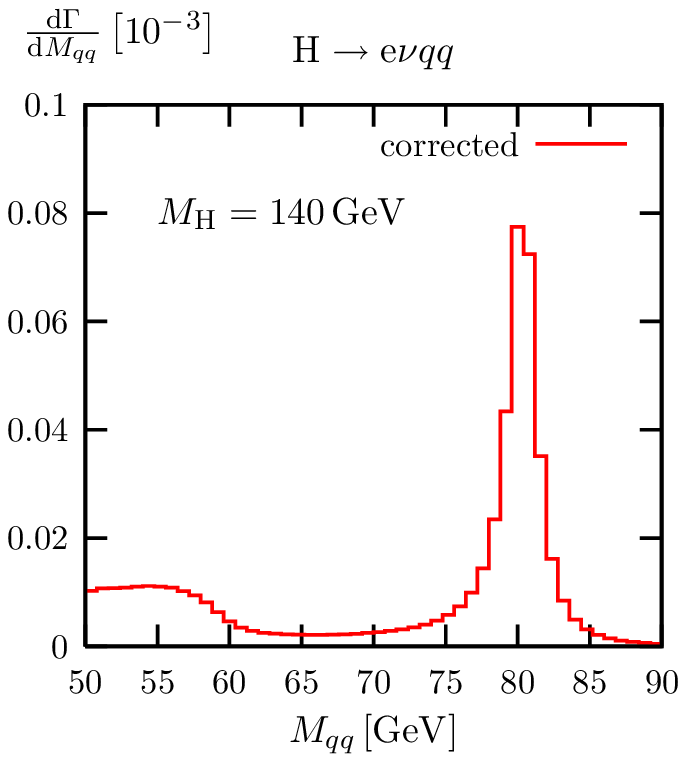}}
\end{picture}
\begin{picture}(7.5,8)
\put(-1.7,-14.5){\includegraphics{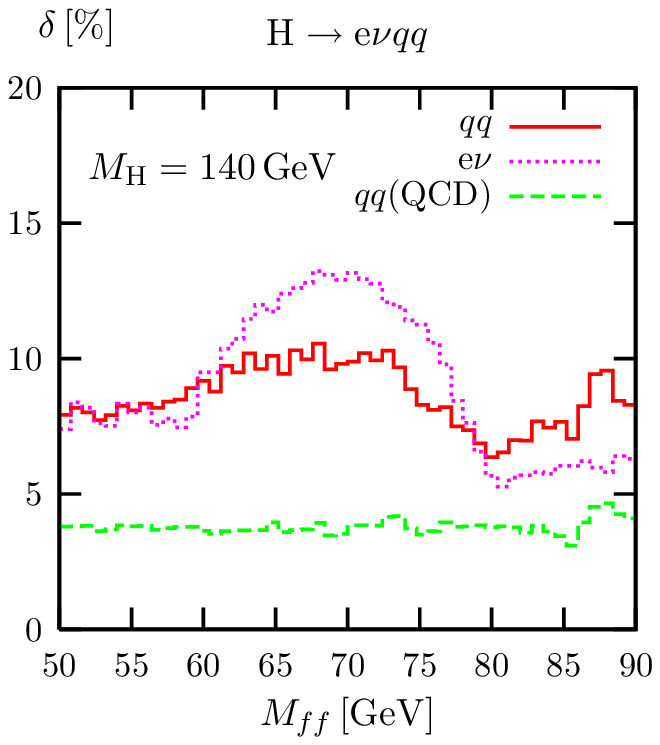}}
\end{picture} }
\centerline{
\begin{picture}(7.7,8)
\put(-1.7,-14.5){\includegraphics{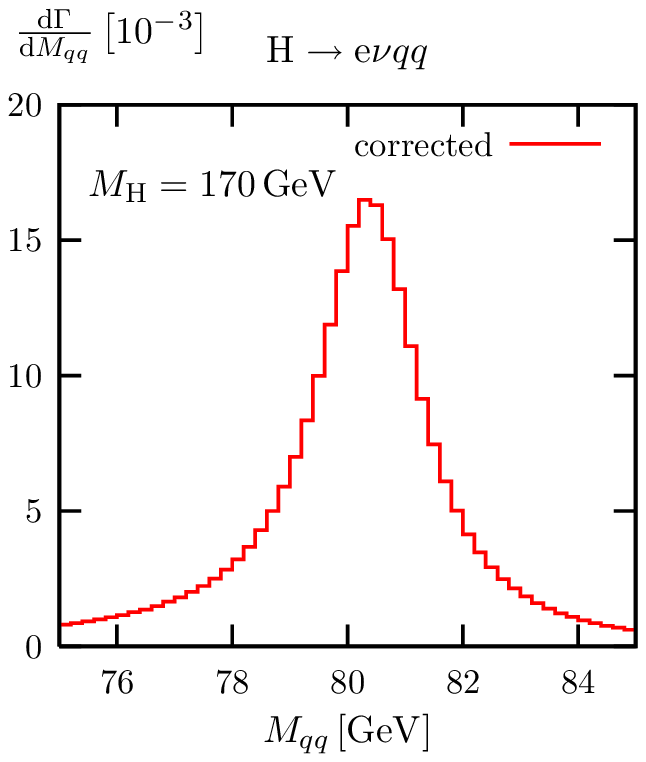}}
\end{picture}
\begin{picture}(7.5,8)
\put(-1.7,-14.5){\includegraphics{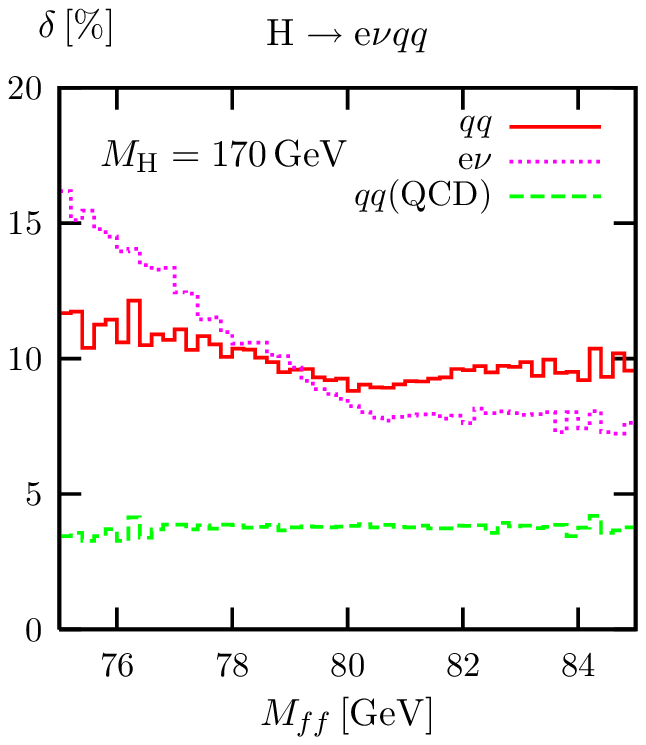}}
\end{picture} }
\caption{Distribution in the invariant mass of the $qq$ pair (l.h.s.) and
  relative EW+QCD corrections to the distributions in the invariant mass of
  the $\mathrm{e}\nu$ and $qq$ pairs (r.h.s) in the decay
  $\PH\to\mathrm{e}\nu qq$ for $\MH=140\GeV$ and $\MH=170\GeV$. For
  the distribution in $M_{qq}$ the relative QCD corrections are
  separately shown.}
\label{fig:enuqqinv}
\end{figure}

In \reffi{fig:enuqqinv} we show the distribution in the invariant mass
of the $qq$ pair and relative corrections to the distributions in the
invariant mass of the $\mathrm{e}\nu$ and $qq$ pairs in the decay
$\PH\to\mathrm{e}\nu qq$ for $\MH=140\GeV$ and $\MH=170\GeV$. 
Similarly to the decay $\PH\to\mathrm{ee}qq$ there is a resonance
around the W-boson mass and, for $\MH<2\MW$, an additional enhancement
for $M_{qq}<\MH-\MW\approx60\GeV$ where the $\Pe\nu$ pair can
become resonant.  Also the relative corrections show the same
characteristics. The corrections for the $M_{\Pe\nu}$ distribution are
somewhat smaller than for the $M_{\Pe\Pe}$ distribution in
\reffi{fig:eeqqinv}, since the neutrino does not radiate photons.

\subsection{Angular distributions}

\begin{figure}
\setlength{\unitlength}{1cm}
\centerline{
\begin{picture}(7.7,8)
\put(-1.7,-14.5){\includegraphics{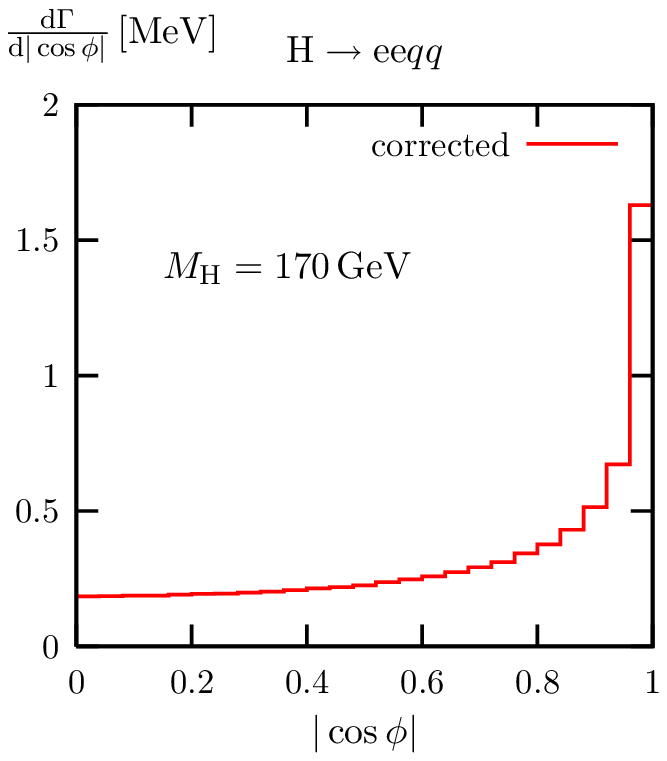}}
\end{picture}
\begin{picture}(7.5,8)
\put(-1.7,-14.5){\includegraphics{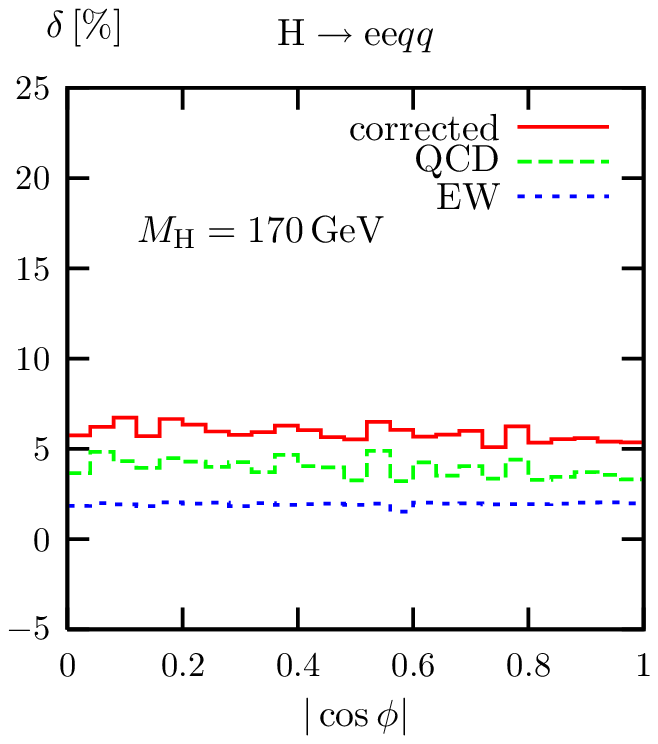}}
\end{picture} }
\centerline{
\begin{picture}(7.7,8)
\put(-1.7,-14.5){\includegraphics{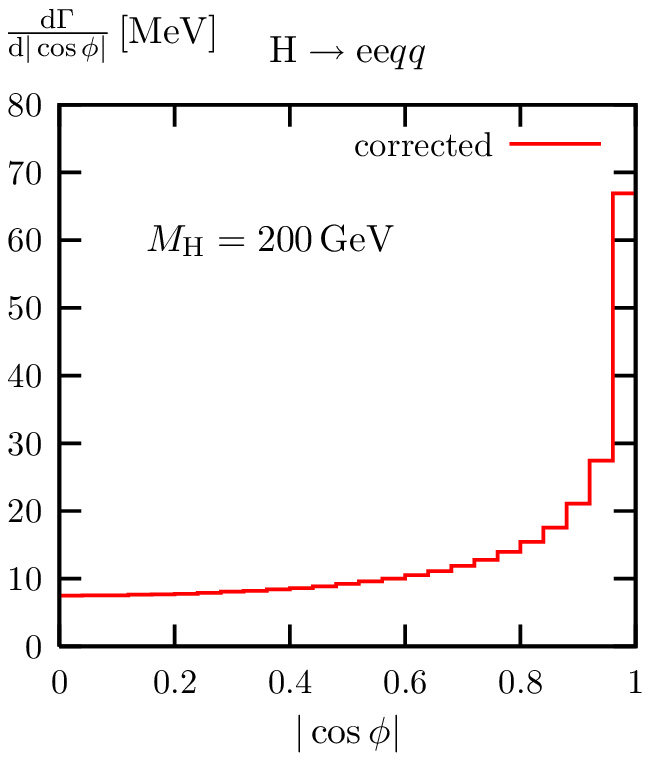}}
\end{picture}
\begin{picture}(7.5,8)
\put(-1.7,-14.5){\includegraphics{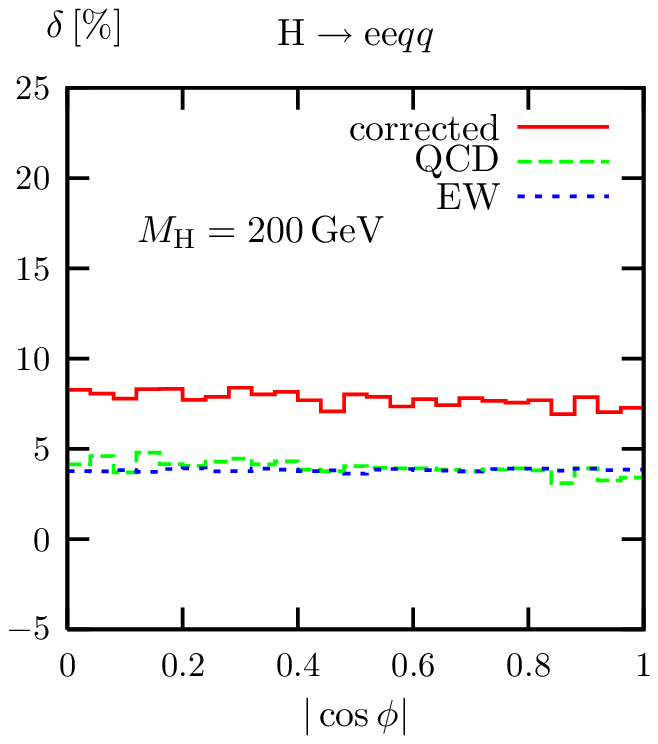}}
\end{picture} }
\caption{Distribution in the angle between the $\PZ\to\mathrm{ee}$ and
  $\PZ\to qq$ decay planes in the decay $\PH\to\mathrm{ee}qq$
  (l.h.s.)\ and corresponding relative 
  EW and QCD corrections (r.h.s.)\ for $\MH=170\GeV$ and
  $\MH=200\GeV$. }
\label{fig:eeqqcosphi}
\end{figure}

Angular distributions can be used to discriminate the Higgs-boson
signal from the background or to study the properties of the Higgs
boson. In \reffi{fig:eeqqcosphi} we show the distribution in the angle
between the decay planes of the reconstructed Z bosons in the decay
$\PH\to\mathrm{ee}qq$ in the rest frame of the Higgs boson.
This angle can, for instance, be used to determine the parity of the
Higgs boson \cite{Nelson:1986ki}.  Since the two jets cannot be
distinguished, we show the distribution in the variable
\beq
|{\cos{\phi}}| = 
\frac{|({\bf k}_{\mathrm{had}}\times{\bf k}_1)({\bf k}_{\mathrm{jet1}}\times{\bf k}_{\mathrm{jet2}})|}
     {|{\bf k}_{\mathrm{had}}\times{\bf k}_1||{\bf k}_{\mathrm{jet1}}\times{\bf k}_{\mathrm{jet2}}|},
\eeq
which is symmetric with respect to the interchange of the jet momenta
${\bf k}_{\mathrm{jet1}}$ and ${\bf k}_{\mathrm{jet2}}$.  Here the
total hadronic momentum ${\bf k}_{\mathrm{had}}$ is equal to the sum
of the two jet momenta, ${\bf k}_{\mathrm{jet1}}+{\bf
  k}_{\mathrm{jet2}}$, because we enforce 2-jet events, and ${\bf
  k}_1$ is the momentum of the electron. For $\MH=200\GeV$, both QCD
and EW corrections are positive and about 4\%.  For $\MH=170\GeV$, the
EW corrections are only about 2\%.  Both the EW and QCD corrections to
this distribution are flat, in contrast to the EW corrections to the
distribution in $\cos\phi^{(\prime)}$ shown in
\citere{Bredenstein:2006rh} for analogous definitions of angles
$\phi^{(\prime)}$ between the two planes defined by leptonically
decaying Z~bosons. This difference results from the fact that the sign
of $\cos\phi^{(\prime)}$ is only observable in the purely leptonic case.

\begin{figure}
\setlength{\unitlength}{1cm}
\centerline{
\begin{picture}(7.7,8)
\put(-1.7,-14.5){\includegraphics{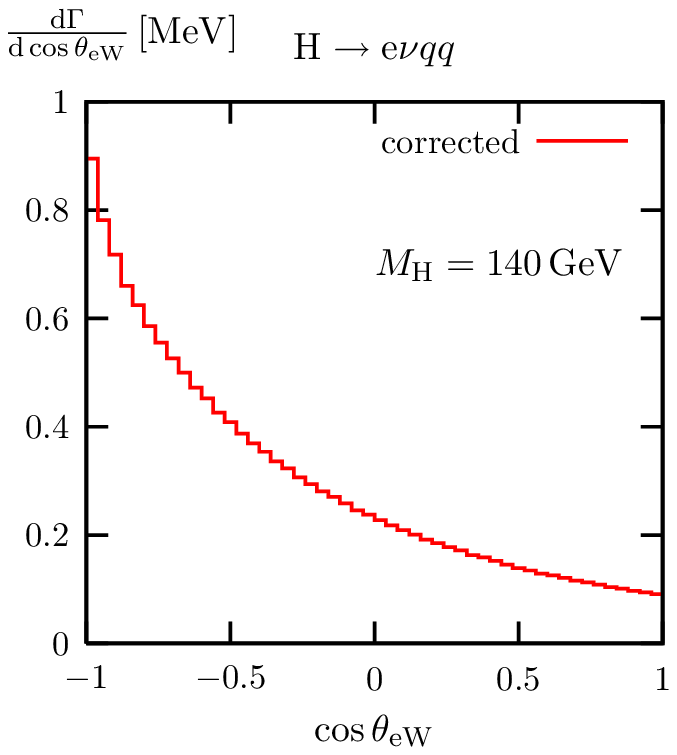}}
\end{picture}
\begin{picture}(7.5,8)
\put(-1.7,-14.5){\includegraphics{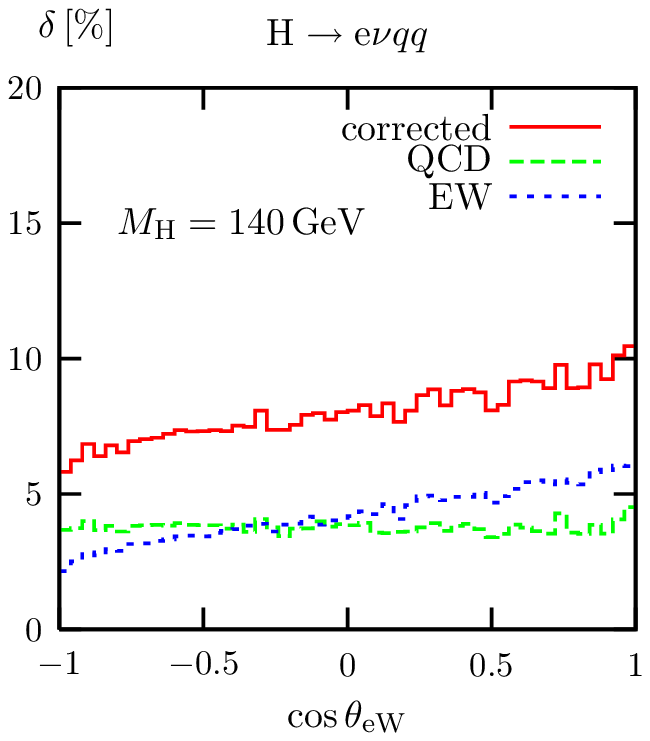}}
\end{picture} }
\centerline{
\begin{picture}(7.7,8)
\put(-1.7,-14.5){\includegraphics{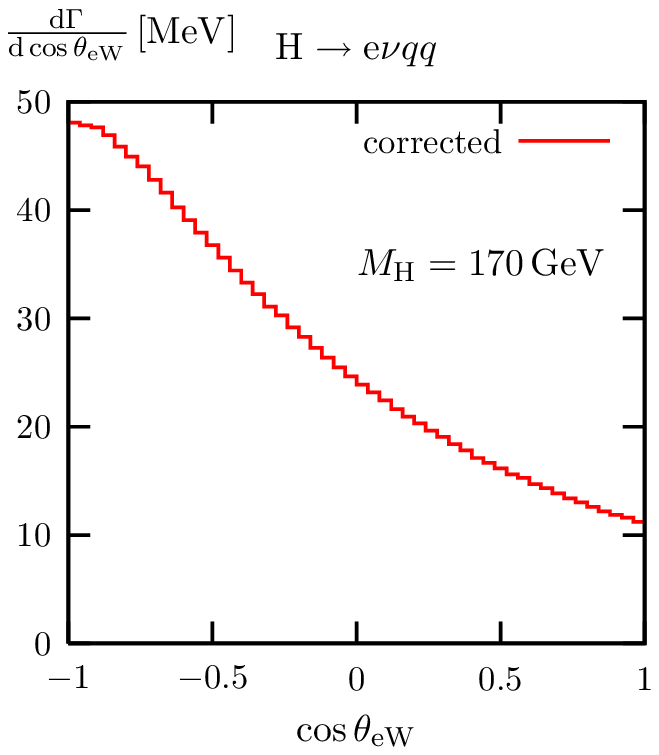}}
\end{picture}
\begin{picture}(7.5,8)
\put(-1.7,-14.5){\includegraphics{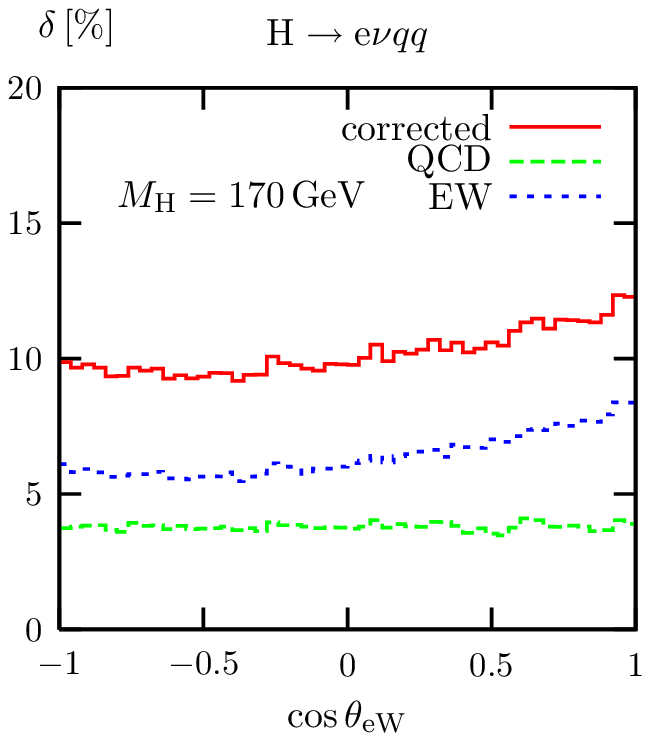}}
\end{picture} }
\caption{Distribution in the angle between the electron and the W~boson
  reconstructed from the $qq$ pair (l.h.s.)\ and corresponding relative
  EW and QCD corrections (r.h.s.)\ 
  in the decay $\PH\to\mathrm{e}\nu qq$ for $\MH=140\GeV$ and
  $\MH=170\GeV$.}
\label{fig:enuqqcthWe}
\end{figure}

In the decay $\PH\to\mathrm{e}\nu qq$, angles between the electron and
jets can be used for background reduction \cite{Cavasinni:2002}.  In
\reffi{fig:enuqqcthWe} we show the distribution in the angle between
the electron and the W boson that is reconstructed from the $qq$ pair
in the rest frame of the Higgs boson and the corresponding relative
QCD and EW corrections.
The plot shows the well-known property that the electron is
predominantly produced in the direction opposite to the hadronically
decaying W~boson.  The QCD corrections are about 4\% and the EW
corrections at the level of 5\%. The complete corrections can reach up
to 12\% depending on the value of the Higgs-boson mass. Since the EW
corrections depend on the angle they distort the distribution by a few
per cent.  
\pagebreak[4]

\section{Conclusions}
\label{se:concl}

The decays of the Standard Model Higgs boson into four fermions via a
W-boson or Z-boson pair lead to experimental signatures at the LHC and
at a future $\Pep\Pem$ linear collider that are both important for the
search for the Higgs boson and for studying its properties. In order
to allow for adequate theoretical predictions for these decays, a
Monte Carlo event generator is needed that properly accounts for the
relevant radiative corrections. {\sc Prophecy4f} is such an event
generator which provides accurate predictions above, in the vicinity
of, and below the WW and ZZ~thresholds, owing to the use of the
complex-mass scheme for the treatment of the gauge-boson resonances.

While {\sc Prophecy4f} originally contained only the electroweak
corrections, in this paper we have included also the complete
$\Ord(\als)$ QCD corrections. This allows to study precise predictions
for all leptonic, semileptonic, and hadronic final states.

The QCD corrections to the partial decay widths are dominated by the
corrections to the gauge-boson decays and roughly given by
$\als/\pi\approx 3.8\%$ for semileptonic and $2\als/\pi\approx 7.6\%$
for hadronic final states. The electroweak corrections to the partial
decay widths are very similar for leptonic, hadronic, and semileptonic
final states. They are positive, typically amount to some per cent,
increase with growing Higgs mass $\MH$, and reach about 8\% at
$\MH\sim500\GeV$. In the on-shell (narrow-width) approximation for the
intermediate gauge bosons, the correction is good within $1$--$2\%$ of
the partial widths for Higgs-boson masses sufficiently above the
corresponding gauge-boson pair threshold, as long as the lowest-order
prediction consistently includes the off-shell effects of the gauge
bosons.  For $\PH\to\PW\PW\to4f$ the narrow-width approximation fails
badly close to the $\PW\PW$ threshold, because the instability of the
W~bosons significantly influences the Coulomb singularity near
threshold.  Only a calculation that keeps the full off-shellness of
the W and Z~bosons can describe the threshold regions properly. A
simple improved Born approximation for the partial widths reproduces
the full calculation within $\lsim2$--$3\%$ for Higgs-boson masses
below $400\GeV$. In this regime our complete calculation should have a
theoretical uncertainty below 1\%. For larger Higgs-boson masses we
expect that unknown two-loop corrections that are enhanced by
$\GF\MH^2$ deteriorate the accuracy.  Finally, for $\MH\gsim700\GeV$
it is well known that perturbative predictions become questionable in
general.

We have numerically investigated distributions for semileptonic final
states where collinear photons are recombined and 2-jet events are
forced.  For angular and invariant-mass distributions the QCD
corrections are flat and reflect the corresponding corrections to the
integrated decay widths.  For angular distributions, which can be used
for background reduction or the study of the quantum numbers of the
Higgs boson, the electroweak corrections are of the order of
$5$--$10\%$ and, in general, distort the shapes.  For invariant-mass
distributions of fermion pairs, which are relevant for the
reconstruction of the gauge bosons, well-known large photonic
corrections show up and can exceed $10\%$ depending on the treatment
of photon radiation.

{\sloppy This work completes the physics part of the Monte Carlo event
  generator {\sc Prophecy4f} for $\PH\to\PW\PW/\PZ\PZ$ $\to4f$. It now
  includes the complete $\Ord(\al)$ electroweak and $\Ord(\als)$ QCD
  corrections as well as corrections beyond $\Ord(\al)$ originating
  from heavy-Higgs effects and final-state photon radiation for all
  possible 4-fermion final states.  {\sc Prophecy4f} works at the
  parton level and generates weighted events; unweighted event
  generation and an interface to parton showering will be addressed in
  the future.  }

\section*{Acknowledgements}
A.B. was supported by a fellowship within the
Postdoc programme of the German Academic Exchange Service (DAAD).

\end{document}